\documentclass[12pt]{article}%
\usepackage[left=2.5cm, right=2.5cm, top=2.25cm, bottom=2.5cm]{geometry}
\usepackage{natbib}
\usepackage{amsfonts}
\usepackage{amssymb}
\usepackage{amsmath}
\usepackage{lscape}
\usepackage{booktabs}
\usepackage{color}
\usepackage{algorithm}
\usepackage[noend]{algpseudocode}
\usepackage{graphicx}
\usepackage{pgfplots}
\usepackage{pgfplotstable}
\usepackage{epstopdf}
\usepackage{subcaption}
\usepackage{caption}
\usepackage{tikz}
\usepackage[normalem]{ulem}%
\providecommand{\U}[1]{\protect\rule{.1in}{.1in}}

\graphicspath{ {./images/} }
\pgfplotsset{compat=1.8}
\usepgfplotslibrary{groupplots}
\pgfplotsset{compat=newest}
\pgfplotsset{plot coordinates/math parser=false}
\newlength\figureheight
\newlength\figurewidth
\pgfplotsset{yticklabel style={text width=3em,align=right}}
\pgfplotsset{xticklabel style={text width=2em,align=right}}

\makeatletter

\def\1{1\!{\rm l}}
\def\BState{\State\hskip-\ALG@thistlm}
\makeatother
\begin{document}

\title{ABC-based Forecasting in State Space Models\thanks{{\footnotesize Weerasinghe
has been supported by a Monash Graduate Scholarship. Martin and Frazier have
been supported by Australian Research Council Discovery Grant DP200101414.
Frazier has also been supported by Australian Research Council Discovery Early
Career Researcher Award DE200101070}, {\footnotesize and Loaiza-Maya supported
by Australian Research Council Discovery Early Career Researcher Award
DE230100029. }}}
\author{Chaya Weerasinghe\thanks{{\footnotesize Corresponding author. Email:
Chaya.W.PatabediMuhamdiramalage@monash.edu}}, Ruben Loaiza-Maya, Gael M.
Martin and David T. Frazier\bigskip\\Department of Econometrics and Business Statistics, Monash University, Australia}
\maketitle

\begin{abstract}
Approximate Bayesian Computation (ABC) has gained popularity as a method for
conducting inference and forecasting in complex models, most notably those
which are intractable in some sense. In this paper we use ABC to produce
probabilistic forecasts in state space models (SSMs). Whilst ABC-based
forecasting in correctly-specified SSMs has been studied, the misspecified
case has not been investigated, and it is that case which we emphasize. We
invoke recent principles of `focused' Bayesian prediction, whereby Bayesian
updates are driven by a scoring rule that rewards predictive accuracy; the aim
being to produce predictives that perform well in that rule, despite
misspecification. Two methods are investigated for producing the focused
predictions. In a simulation setting, `coherent' predictions are in evidence
for both methods: the predictive constructed via the use of a particular
scoring rule predicts best according to that rule. Importantly, both focused
methods typically produce more accurate forecasts than an exact, but
misspecified, predictive. An empirical application to a truly intractable SSM
completes the paper.

\medskip


\end{abstract}

\smallskip

\noindent\emph{MSC2010 Subject Classification}: 62E17, 62F15, 62M20 \smallskip



\newpage

\section{Introduction}

State space models (SSMs) are a key forecasting tool used in a wide range of
areas including astronomy, ecology, economics and finance. Whilst the
predictive distributions for such models are often readily accessible via
Bayesian Markov chain Monte Carlo (MCMC) methods (see \citealp{giordani2011},
and \citealp{fearnhead2011}, for reviews) or particle MCMC variants thereof
(\citealp{andrieu:doucet:holenstein:2010};
\citealp{Flury2011}){\normalsize {{,}}} challenges remain when the model is
`intractable' in some sense. Intractable SSMs come in two varieties:
\textit{First}, those for which the dimension of either the data or the latent
states, or both, is very large; \textit{Second}, those for which some
component of the model does not admit an analytical representation as a
probability mass, or density function. Exact Bayesian methods based on (P)MCMC
are computationally impracticable in the first case, and typically infeasible
in the second.

High-dimensional SSMs are increasingly tackled via variational methods
(\citealp{blei2017variational}; \citealp{zhang2018advances}), which are able
to produce approximations to both the posterior and predictive distributions
in a reasonable computing time (\citealp{tran2017variational};
\citealp{koop2018variational}; \citealp{quiroz2018gaussian};
\citealp{chan2020fast}; \citealp{loaiza2020fast};
\citealp{frazier2023variational}). SSMs with unavailable components, on the
other hand, have been managed via approximate Bayesian computation (ABC)
(\citealp{dean:singh:jasra:peters:2011}; \citealp{CREEL2015};
\citealp{frazier2019approximate}; \citealp{martin2019auxiliary}). This paper
continues in the latter vein, by looking explicitly at the use of ABC to
conduct \textit{forecasting} in SSMs.

Whilst ABC-based forecasts for SSMs were first systematically studied in
\cite{frazier2019approximate}, the emphasis therein was entirely on the
correctly-specified case; with exploration of the performance of the forecasts
under misspecification of the data generating process (DGP) left as a future
goal. It is that goal that we now pursue as part of this paper. Moreover, for
the first time, we meld the principles of ABC-based forecasting with recent
developments in so-called `generalized', `loss-based', or `focused' Bayesian
prediction, in which performance according to a user-specified measure of
predictive accuracy is the goal, rather than correct model specification
\textit{per se }(\citealp{loaiza2021focused} and
\citealp{frazierloss})\textit{.} In so doing, we show that, not only does ABC
provide a way of managing \textit{intractable } models (i.e., the usual
motivation for ABC), it is also a feasible way of implementing loss-based
prediction \textit{per se} in the state space setting, due to the
computational difficulty of specifying `exactly' a loss-based criterion in an SSM.


ABC constructs an approximation to an exact posterior via simulation from the
assumed model, avoiding specification of the (unavailable) likelihood function
altogether. In the simplest scenario, draws of the unknown parameters from the
prior that yield simulated data that `match' the observed data -- through the
prism of selected summary statistics -- are used to produce the approximation
of the exact posterior; the accuracy of the approximation depending, amongst
other things, on the closeness of the summaries to sufficiency. Under the
condition that the assumed model correctly specifies the true DGP, and under
appropriate regularity, the ABC posterior has been shown to be Bayesian
consistent for the true parameter (vector), to be asymptotically normal, and
to produce a posterior mean with a sampling distribution that is
asymptotically normal (\citealp{frazier2018asymptotic}). In an explicitly
state space setting \cite{martin2019auxiliary} also prove the consistency of
the ABC posterior for the static (or `global') parameters. Moreover,
\cite{frazier2019approximate} have shown that -- again, under correct
specification -- ABC-based forecasting (which they term `approximate Bayesian
forecasting' (ABF)) yields equivalent results in large samples to exact
Bayesian forecasting, with both the approximate and exact predictives
`merging', in turn, with the true predictive distribution. Once again, this
result nests the case of prediction in an SSM.

Critically, as emphasized, all of the ABC work cited above operates explicitly
in a correctly-specified scenario. In contrast, \cite{frazier2020model}
explore the properties of ABC inference under misspecification. The authors
demonstrate that while the ABC posterior does concentrate onto an
appropriately defined pseudo-true value in the limit, it possesses nonstandard
asymptotic behaviour. The implications of these results for ABC-based
forecasting are left unexplored -- and that is the key focus of the current
paper, specifically in the context of misspecified \textit{SSMs}. However,
once the assumption of correct model specification is no longer maintained,
the whole tenor of the Bayesian forecasting exercise needs to change to one in
which focus is directed at the particular goal of the forecasting exercise.
Such is the argument put forward in the recent forecasting work by
\cite{loaiza2021focused} and \cite{frazierloss}, and is the one that we pursue here.

In short, the aim is to replace the (potentially misspecified) likelihood
function in the conventional Bayesian up-date with a criterion function that
rewards a user-specified form of forecast accuracy or, equivalently, penalizes
the associated forecast loss. In the current context, in which inference is
conducted via ABC, the approach involves selecting summary statistics that
give weight to ABC draws that will -- in turn -- lead to predictive accuracy
in a selected scoring rule. Accordingly, we refer to this process as
loss-based approximate Bayesian forecasting, or `loss-based ABF' for short. We
implement this approach by embedding within the ABC algorithm a loss-based
criterion function defined in terms of an `auxiliary model'
(\citealp{drovandi2011approximate}; \citealp{drovandi2015bayesian};
\citealp{martin2019auxiliary}) that is a good match to the assumed SSM, but
which has a closed-form predictive. The sample criterion is built from the
scoring rule of interest, with maximization of that criterion producing the
summary statistics used to drive the ABC algorithm. We compare this approach
with the alternative of using the auxiliary model \textit{itself} (i.e.
independent of the ABC mechanism) to produce suitably focused predictive
distributions, via a generalized Bayesian up-date.

The paper proceeds as follows. In Section 2 we provide the appropriate
background for the forecasting approach advocated herein, by outlining the key
aspects of performing both exact and ABC-based forecasting in
\textit{correctly-}specified SSMs. In Section 3 we then illustrate the
proposed loss-based ABF approach in a misspecified setting. As a comparator,
as noted above, the auxiliary model adopted within the ABC algorithm is used,
in its own right, to produce a loss-based predictive \textit{directly }via a
generalized Bayesian up-date (without an ABC step). To keep the nomenclature
clear, we refer to this second approach as focused Bayesian prediction (or
`FBP') given the sense in which it directly mimics the FBP approach proposed
in \cite{loaiza2021focused}. Section 4 then illustrates the performance of
both loss-based ABF and FBP in an extensive set of simulation experiments,
under both a correctly-specified and misspecified true SSM. This includes
comparison with the exact likelihood-based predictive, which is accessible in
this artificial setting.\textit{ }The key result is that, under model
misspecification, both loss-based ABF and FBP produce coherent predictions.
That is, using an approach with a particular focus yields the best
out-of-sample performance according to that same measure of predictive
accuracy. Moreover, the predictions produced by both focused methods are more
often than not superior to those of the (misspecified) exact predictive. When
comparing the relative performance of loss-based ABF and FBP the conclusions
are mixed; however the former approach often achieves better accuracy than the
latter. Section 5 then applies both predictive methods to a stochastic
volatility model with an intractable $\alpha-$ stable transition. Coherence is
still a feature of both focused methods. However, in this case loss-based ABF
produces average scores out-of-sample that are larger, overall, than those of
FBP, suggesting that there are indeed benefits of performing the focussing
within an ABC algorithm that is driven - in part - by a well specified SSM.
The paper concludes in Section 6.

\section{Bayesian Forecasting in Correctly-Specified SSMs\label{sec2}}

\subsection{Exact Bayesian forecasting\label{sec2.1}}

We assume a stationary ergodic process $\{Y_{t}\}_{t\geq0}$ taking values in a
measure space $(\mathsf{Y},\mathcal{F}_{y})$, with $\mathcal{F}_{y}$ a Borel
$\sigma$-field, specified according to an SSM that depends on an unobserved
state process $\{X_{t}\}_{t\geq0}$, taking values in a measure space
$(\mathsf{X},\mathcal{F}_{x})$, with $\mathcal{F}_{x}$ a Borel $\sigma$-field.
Conditional on $\{X_{t}\}$, the sequence $\{Y_{t}\}$ is independent. {To
simplify the exposition we choose to illustrate our approach in the case where
both }$X_{t}$ {and }$Y_{t}$ {are continuous scalar random variables, with
values }$x_{t},$ $y_{t}$, $t=1,...,,T${ and initial state }$x_{1}.$ For $t>0$,
we assume the following measurement and transition densities:
\begin{equation}
p(y_{t}|x_{t},\theta) \label{meas}%
\end{equation}%
\begin{equation}
p(x_{t}|x_{t-1},\theta), \label{state}%
\end{equation}
with an initial state density,%
\begin{equation}
p(x_{1}|\theta), \label{initial}%
\end{equation}
where $\theta$ is a $P-$dimensional vector of unknown parameters. {Whilst
extension to the multivariate case (for either }$X_{t}$ or $Y_{t}$), with a
concurrent increase in the dimension of $\theta$, would cause no additional
conceptual problems, ABC is not inherently well-suited to settings where
$\theta$ is high-dimensional (\citealp{martin2023approximating}). The
first-order Markovian assumption for $X_{t}$ is innocuous, and any finite (and
known) Markov order can be accommodated by an appropriate definition of
$X_{t}$.

For $p(\theta)$ defining the prior density, the posterior for $\theta$ can be
expressed as
\begin{equation}
p(\theta|y_{1:T})=\int_{\mathsf{X}}p(\theta,x_{1:T}|y_{1:T})\,dx_{1}\dots
dx_{T}, \label{exact_post}%
\end{equation}
where $p(\theta,x_{1:T}|y_{1:T})\propto p(y_{1:T}|x_{1:T},\theta
)p(x_{1:T}|\theta)p(\theta)$, with $x_{1:T}=(x_{1},...,x_{T})^{\prime}$ and
$y_{1:T}=(y_{1},...,y_{T})^{\prime}$. In certain special cases (e.g. when
(\ref{meas}) to (\ref{initial}) defines a linear Gaussian SSM), a Gibbs
sampling scheme can be used to produce draws from $p(\theta,x_{1:T}|y_{1:T})$
and, thus, from $p(\theta|y_{1:T})$ (\citealp{carter:kohn:1994};
\citealp{fruhwirth-schnatter:1994}). Typically, however, the conditionals
$p(\theta|x_{1:T},y_{1:T})$ and $p(x_{1:T}|\theta,y_{1:T})$ will not have
known closed forms, and either a Metropolis Hasting (MH)-within-Gibbs scheme
(\citealp{jacquier94}; \citealp{kim1998svl}; \citealp{stroud2003};
\citealp{STRICKLAND2006}), or a PMCMC algorithm
(\citealp{andrieu:doucet:holenstein:2010}; \citealp{Flury2011}) is applied.
(See \citealp{giordani2011} and \citealp{jacquier2011bayesian} for reviews.)
Once draws of $\theta$ and $x_{1:T}$ have been produced from $p(\theta
,x_{1:T}|y_{1:T})$, the predictive probability density function (pdf),%
\begin{equation}
p(y_{T+1}|y_{1:T})=\int_{\mathsf{X}}\int_{\mathsf{X}}\int_{{\Theta}}%
p(y_{T+1}|x_{T+1},\theta)p(x_{T+1}|x_{T},\theta)p(\theta,x_{1:T}%
|y_{1:T})d\theta dx_{1:T}d{x}_{T+1}, \label{forecast_ss}%
\end{equation}
can be estimated using kernel density estimation methods, using subsequent
draws from the transition and measurement densities, $p(x_{T+1}|x_{T},\theta)$
and $p(y_{T+1}|x_{T+1},\theta)$ respectively, or by averaging the measurement
densities over the draws of $x_{T+1}$ and $\theta$.

However, this approach is operational only if both components of the
(complete) likelihood function, $p(y_{1:T}|x_{1:T},\theta)p(x_{1:T}|\theta)$,
are\ available in closed form (for an MH-within-Gibbs scheme) or an unbiased
estimator of each is available via particle filtering (for a PMCMC scheme). If
neither of these conditions hold, draws from the augmented posterior,
$p(\theta,x_{1:T}|y_{1:T})$, are unavailable and an estimate of $p(y_{T+1}%
|y_{1:T})$ that is exact up to simulation error cannot be produced. This is
where ABC, and the associated production of an \textit{approximation }of
$p(y_{T+1}|y_{1:T})$, come into play.

\subsection{Approximate Bayesian forecasting (ABF)}

The aim of ABC is to produce draws from an approximation to $p(\theta
|y_{1:T})$ in the case where $p(y_{1:T}|x_{1:T},\theta)$, $p(x_{1:T}|\theta)$
(and $p(\theta)$) can be at least simulated from, even if either
$p(y_{1:T}|x_{1:T},\theta)$ or $p(x_{1:T}|\theta)$ is unavailable in closed
form. The simplest (accept/reject) form of the algorithm
(\citealp{tavare1997inferring}; \citealp{pritchard1999population}), adapted
for the SSM, proceeds as per Algorithm 1. In Algorithm 1, $\eta(.)$ is a
(vector) statistic, $d\{.\}$ is a distance criterion, and, given $N$, the
tolerance level $\varepsilon$ is chosen to be small.

\begin{algorithm}
\caption{ABC accept/reject algorithm}\label{ABC}
\begin{algorithmic}[1]
\State Simulate $\mathbf{\theta }^{i}$, $i=1,2,...,N$, from $p(\mathbf{\theta }).$
\State Simulate the artificial data $z_{1:T}^{i}=\big(  x_{1:T}^{s}(\theta^{i}), y_{1:T}^{s}(\theta^{i})\big)$ as follows:
	\begin{itemize}
		\item Simulate the states $x_{1:T}^{s}(\theta^{i})$, $i=1,2,...,N$ from $p(x_{1}^{s}| \theta^{i})$ and $p(x_{t}^{s} |x_{t-1}^{s}, \theta^{i})$, for $t=2,...,T$.
		\item Simulate $y_{1:T}^{s}(\theta^{i})$, $i=1,2,...,N$ from $p \big(y_{t}^{s}|x_{t}^{s}(\theta^{i})\big)$, for $t=1,...,T$.
	\end{itemize}
\State Select $\mathbf{\theta }^{i}$ such that:%
\begin{equation}
d\{\mathbf{\eta }(y_{1:T}),\mathbf{\eta }\big( y_{1:T}^{s}(\theta^{i})\big)\}\leq
\varepsilon ,  \label{distance}
\end{equation}%
\end{algorithmic}
\end{algorithm}

Denoting by $x_{1:T}^{s}$ and $y_{1:T}^{s}$, simulated draws from
$p(x_{1:T}|\theta^{i})$ and $p(y_{1:T}|x_{1:T},\theta^{i})$ respectively, an
accepted draw of $\theta$ from Algorithm 1 is a draw from:%
\[
p_{\varepsilon}(\theta|\eta(y_{1:T}))=\frac{\int_{y_{1:T}^{s}}\int%
_{x_{1:T}^{s}}\mathbb{I}_{\varepsilon}[d\{\eta\left(  y_{1:T}^{s}\right)
\mathbf{,}\eta\left(  y_{1:T}\right)  \}\leq\varepsilon]p(y_{1:T}^{s}%
|x_{1:T}^{s},\theta)p(x_{1:T}^{s}\mathbf{|}\theta)p(\theta)dx_{1:T}%
^{s}dy_{1:T}^{s}}{\int_{\Theta}\int_{y_{1:T}^{s}}\int_{x_{1:T}^{s}}%
\mathbb{I}_{\varepsilon}[d\{\eta\left(  y_{1:T}^{s}\right)  \mathbf{,}%
\eta\left(  y_{1:T}\right)  \}\leq\varepsilon]p(y_{1:T}^{s}|x_{1:T}^{s}%
,\theta)p(x_{1:T}^{s}\mathbf{|}\theta)p(\theta)dx_{1:T}^{s}dy_{1:T}^{s}%
d\theta},
\]
where $\mathbb{I}_{\varepsilon}[d\{\eta\left(  y_{1:T}^{s}\right)
\mathbf{,}\eta\left(  y_{1:T}\right)  \}\leq\varepsilon]$ is one if
$d\{\eta\left(  y_{1:T}^{s}\right)  \mathbf{,}\eta\left(  y_{1:T}\right)
\}\leq\varepsilon$ and zero otherwise. When $\eta(\cdot)$ is sufficient for
$\theta$ and for $\varepsilon\rightarrow0$, $p_{\varepsilon}(\theta
|\eta(y_{1:T}))$ is equivalent to the exact posterior, $p(\theta|y_{1:T}).$
Clearly, the sorts of intractable problems to which ABC is applied preclude
sufficiency, almost by default; hence ABC is only ever intrinsically
approximate, even in cases where the computing budget allows the tolerance to
be very small.\footnote{As is usual in practice, we apply a modified version
of Algorithm 1, whereby we replace the acceptance step in Algorithm 1 with a
nearest-neighbour selection step (\citealp{biau2015new}). The accepted draws
of $\theta$ in this version of the algorithm are associated with an empirical
quantile over the simulated distances $d\{\eta\left(  y_{1:T}^{s}\right)
\mathbf{,}\eta\left(  y_{1:T}\right)  \}$. That is, Step 3 in Algorithm 1 is
replaced with the following step: Select all $\theta^{i}$ associated with the
$q=\delta/N$ smallest distances $d\{\eta\left(  y_{1:T}^{s}\right)
\mathbf{,}\eta\left(  y_{1:T}\right)  \}$ for some $\delta$. We refer the
reader to \cite{martin2023approximating} for an outline of the many further
adaptations of the simple accept/reject ABC algorithm that have been proposed,
and for extensive referencing of the ABC literature.}

Exploiting the Markov property of the state process in (\ref{state}), the
associated approximate predictive can be expressed as%

\begin{align}
&  g(y_{T+1}|y_{1:T})\label{approx_pred}\\
&  =\int_{\mathsf{X}}\int_{\mathsf{X}}\int_{{\Theta}}p(y_{T+1}|x_{T+1}%
,\theta)p(x_{T+1}|x_{T},\theta)p(x_{T}|\theta,y_{1:T})p_{\varepsilon}%
(\theta|\eta(y_{1:T}))d\theta dx_{T}d{x}_{T+1}.\nonumber
\end{align}
Given draws of $\theta$ from\textbf{ }$p_{\varepsilon}(\theta|\eta(y_{1:T}))$
via Algorithm 1, all that is needed is a forward-filtering algorithm in order
to draw from\textbf{ }$p(x_{T}|\theta,y_{1:T})$\textbf{ }and to produce a
simulation-based estimate of $g(y_{T+1}|y_{1:T})$, either as a sample mean of
the measurement densities defined by the draws of\textbf{ }$\theta$\textbf{
}(and\textbf{ }$x_{T+1}$), or by applying kernel density techniques using
subsequent draws from the transition and measurement densities, $p(x_{T+1}%
|x_{T},\theta)$ and $p(y_{T+1}|x_{T+1},\theta)$. The key thing here is that
posterior draws from the conditional posterior for the full vector of states
is not required when it comes to implementing ABF. Note also that draws of
$x_{T}$ produced from Algorithm 1, which would implicitly be conditioned on
$\eta\left(  y_{1:T}\right)  $, not $y_{1:T}$, are not used. The particle
filter enables draws of $x_{T}$ to be conditioned on the complete data set,
and this is critical for the resultant accuracy of $g(y_{T+1}|y_{1:T})$ as an
approximation of $p(y_{T+1}|y_{1:T}).$ (See \citealp{frazier2019approximate},
for further discussion.)

Given the obvious importance of the choice of summary statistics in
determining how close $p_{\varepsilon}(\theta|\eta(y_{1:T}))$ can \textit{ever
}be to $p(\theta|y_{1:T})$, attention has been given to maximizing the
information content of the summaries in some way
(\citealp{joyce2008approximately}; \citealp{wegmann2009efficient};
\citealp{blum2010approximate}; \citealp{fearnhead2012constructing}), including
via the maximization of the likelihood function of an auxiliary model. It is
the auxiliary-model approach to ABC that we pursue in this paper, given its
superior performance in the SSM setting (\citealp{martin2019auxiliary}), plus
the ease with which this method allows us to focus the ABC draws in the manner
required in a loss-based setting. We will illustrate this latter point in
detail below, once we have set the scene for the implementation of loss-based
prediction \textit{per se }in the case where the assumed SSM in (\ref{meas})
to (\ref{initial}) is a misspecified representation of the true DGP.

\section{Bayesian Forecasting in Misspecified SSMs}

\subsection{Loss-based Bayesian prediction}

The predictive distribution in (\ref{forecast_ss}), when accessed via an exact
simulation method, is the gold standard in Bayesian prediction, and the
benchmark against which any approximation -- such as that in
(\ref{approx_pred}) -- would be judged. However, its reliability as a
benchmark depends critically on the assumption that the predictive model in
(\ref{meas}) to (\ref{initial}) is correctly specified. The accuracy of the
approximate predictive in (\ref{approx_pred}) is also predicated on the
assumption that the generating model used both in the ABC algorithm, and the
subsequent particle filtering algorithm, tallies with the true data generating
process (DGP) (\citealp{martin2019auxiliary}). Once that assumption is
violated, `all bets are off', and the usefulness -- or otherwise -- of both
(\ref{forecast_ss}) and (\ref{approx_pred}) is solely a function of
\textit{how }misspecified the assumed model is.

In this case, it makes sense to take a different approach to prediction,
namely to seek the form of predictive accuracy that actually matters for the
problem at hand, rather than seeking correct model specification \textit{per
se}. More specifically, a more sensible predictive paradigm involves replacing
the logarithmic scoring rule that explicitly underpins (\ref{forecast_ss}),
and that is implicit in the generative model used in (\ref{approx_pred}), by
the \textit{particular} scoring rule (and associated form of predictive
accuracy) that matters for the particular forecasting problem being tackled.
Such is the thinking that underpins the predictive methodology in
\cite{lacoste2011approximate}, \cite{loaiza2021focused} and \cite{frazierloss}%
, and which we also adopt here.

We begin by assuming a class of plausible predictive SSMs for $Y_{T+1}$,
conditioned on the information $\mathcal{F}_{T}$, indexed by the global
parameter vector $\theta$: $\mathcal{P}^{(T)}:=\{P_{\theta}^{(T)},\theta
\in\Theta\}$. This class may comprise a single parametric SSM, in which
$\theta$ retains its usual interpretation as the vector of parameters that
underpins both the measurement and state distributions, or may comprise some
combination of plausible models, in which case $\theta$ would comprise both
the model-specific parameters and the combination parameters. To keep the
scope of the paper manageable we consider only the case of a class of single
parametric models, and we continue to denote the measurement and transition
densities of the assumed SSM using the notation in (\ref{meas}) to
(\ref{initial}). The key thing is that we now no longer assume that the
\textit{true} predictive distribution is an element of $\mathcal{P}^{(T)}$.

Given the predictive class $\mathcal{P}^{(T)}$, the accuracy of $P_{\theta
}^{(T)}\in\mathcal{P}^{(T)}$ can be measured using the positively-oriented
proper scoring rule $s:\mathcal{P}^{(n)}\times\mathcal{Y}\mapsto\mathbb{R}$,
{where the expected scoring rule under the true distribution }$P_{0}${\ is
defined as}
\begin{equation}
\mathbb{S}(\cdot,P_{0}):=\int_{y\in\Omega}s(\cdot,y)dP_{0}(y).
\label{exp_score}%
\end{equation}
(See \citealp{gneiting2007strictly}, for details of proper scoring rules).
{Since} $\mathbb{S}(\cdot,P_{0})$ cannot be attained in practice, a sample
estimate based on $y_{1:T}$ is used to define the sample criterion, for a
given $\theta\in\Theta$, as
\begin{equation}
S_{T}(\theta):=\sum_{t=1}^{T}s(P_{\theta}^{(t)},y_{t+1}). \label{sample_crit}%
\end{equation}
Adopting the \textit{generalized }updating rule proposed by
\cite{bissiri2016general}, amongst others, and first used by
\cite{loaiza2021focused} and \cite{frazierloss} in a \textit{forecasting
}setting, we define
\begin{equation}
p_{L}(\theta|y_{1:T})=\frac{\exp\left[  wS_{T}\left(  \theta\right)  \right]
\pi(\theta)}{\int_{\Theta}\exp\left[  wS_{T}\left(  \theta\right)  \right]
\pi(\theta)d\theta}, \label{post}%
\end{equation}
where $w$ is a scale factor, which needs to be set using certain criteria; see
\cite{loaiza2021focused} for details, and relevant earlier references cited
therein. In the spirit of the early work of {{\cite{zhang2006information} and}
{\cite{jiang2008gibbs} }}this posterior may be\textbf{\ }referred to a
\textit{Gibbs} posterior. Alternatively, it may be termed a `focused'
posterior following \cite{loaiza2021focused}. Given that a negatively-oriented
scoring rule can be viewed as a measure of loss in a predictive setting, we
default to the term `loss-based' in the main, and use the subscript $L$ in
(\ref{post}) to signal this nomenclature.

By design, (\ref{post}) places higher weight on elements of $\Theta$ that lead
to predictive models, $P_{\theta}^{(T)}$, with higher predictive accuracy in
the scoring rule $s(\cdot,\cdot)$ (or, equivalently, to models with lower
predictive loss). As a result, the predictive defined by averaging with
respect to $p_{L}(\theta|y_{1:T})$ rather than $p(\theta|y_{1:T})$ in
(\ref{exact_post}):
\begin{align}
&  p_{L}(y_{T+1}|y_{1:T})\label{Gibbs_pred}\\
&  =\int_{\mathsf{X}}\int_{\mathsf{X}}\int_{{\Theta}}p(y_{T+1}|x_{T+1}%
,\theta)p(x_{T+1}|x_{T},\theta)p(x_{T}|\theta,y_{1:T})p_{L}(\theta
|y_{1:T})d\theta dx_{T}d{x}_{T+1},\nonumber
\end{align}
may well outperform, in the chosen rule $s(\cdot,\cdot)$, the predictive in
(\ref{forecast_ss}) constructed using the exact -- but misspecified \ --
posterior in (\ref{exact_post}). Given its explicit dependence on the
loss-based posterior, the predictive in (\ref{Gibbs_pred}) is referred to as
the loss-based predictive (with the subscript $L$ signalling this
terminology), and with the additional adjective `exact' also used whenever
$p_{L}(y_{T+1}|y_{1:T})$ is accessed by an exact sampling scheme.

Whilst the superior predictive performance of an exact loss-based predictive
has been established -- both theoretically and numerically -- in
\cite{loaiza2021focused}, and similar validity demonstrated in
\cite{frazierloss} for the predictive based on a variational Bayes
approximation to $p_{L}(\theta|y_{1:T})$, both papers assume only
\textit{observation-driven} predictive classes. Our focus here is on the case
where $\mathcal{P}^{(T)}$ is defined in terms of SSMs; and, as is clear from
inspection of (\ref{sample_crit}), this causes an immediate problem.
Specification of the sum of scores requires $P_{\theta}^{(t)}$ to be expressed
as a function of $\theta$ alone, in order for the `marginal' posterior,
$p_{L}(\theta|y_{1:T})$, to be produced. This, in turn, requires the states to
be integrated out. Whilst this could, in principle, be achieved in a
preliminary step via numerical means, this is impractical from a computational
point of view. Moreover, once the scenario of a truly intractable problem is
entertained due to some component of the assumed model being unavailable
analytically, $p_{L}(\theta|y_{1:T})$ is essentially out of reach.
\textit{Both }of these problems can be solved by the use of ABC.

\subsection{Loss-based ABF\label{loss-based ABF}}

In the spirit of \cite{drovandi2011approximate}, \cite{drovandi2015bayesian}
and \cite{martin2019auxiliary} we implement an ABC approach by producing a
vector of summary statistics, $\eta(\cdot)$, by \textit{maximizing} a sample
criterion function of the form of (\ref{sample_crit}), but defined in terms of
an auxiliary model chosen to be a reasonable representation of the assumed SSM
in (\ref{meas}) to (\ref{initial}). A key assumption in our choice of
auxiliary model is that it admits a closed-form predictive. This allows for
the specification of a score-based criterion as:%
\begin{equation}
S_{T}(\beta):=\sum_{t=1}^{T}s(P_{\beta}^{(t)},y_{t+1}), \label{aux_crit}%
\end{equation}
where $P_{\beta}^{(t)}$ denotes the conditional predictive associated with the
auxiliary model, with parameter vector $\beta$. To keep the analysis
manageable, we use three alternative forms of (positively-oriented) scoring
rules:
%

\begin{align}
&  s_{LS}\left(  P_{\beta}^{(t)},y_{t+1}\right)  =\log p_{\beta}%
(y_{t+1}|y_{1:t}),\label{ls}\\
&  s_{CRPS}\left(  P_{\beta}^{(t)},y_{t+1}\right)  =-\int_{-\infty}^{\infty
}\left[  P_{\beta}^{(t)}-I(y\geq y_{t+1})\right]  ^{2}dy,\label{crps}\\
&  s_{CLS}\left(  P_{\beta}^{(t)},y_{t+1}\right)  =\log p_{\beta}%
(y_{t+1}|y_{1:t})I\left(  y_{t+1}\in A\right)  +\left[  \ln\int_{A^{c}%
}p_{\beta}(y|y_{1:t})dy\right]  I\left(  y_{t+1}\in A^{c}\right)
,\label{cls}\\
&  s_{\text{IS}}\left(  P_{\beta}^{(t)},y_{t+1}\right)  =\left(
u_{t+1}-l_{t+1}\right)  +\frac{2}{\alpha}\left(  l_{t+1}-y_{t+1}\right)
I\left(  y_{t+1}<l_{t+1}\right)  +\frac{2}{\alpha}\left(  y_{t+1}%
-u_{t+1}\right)  I\left(  y_{t+1}>u_{t+1}\right)  , \label{msis}%
\end{align}
where $p_{\beta}(y_{t+1}|y_{1:t})$ represents the predictive density function
associated with $P_{\beta}^{(t)}$, evaluated at the observed\textbf{\ }%
$y_{t+1}$. The log-score (LS) in (\ref{ls}) rewards a high value if the
observed value, $y_{t+1}$ is in high density region of $p_{\beta}(.|y_{1:T})$,
and maximization of (\ref{aux_crit}) defined using the log-score yields the
maximum likelihood estimator (MLE) of $\beta.$ The continuously ranked
probability score (CRPS) in (\ref{crps}) (see \citealp{gneiting2007strictly})
is sensitive to distance and rewards the assignment of high predictive mass
near to the realized $y_{t+1}$, rather than just at that value. The censored
log-score (CLS) in (\ref{cls}), first proposed by \cite{diks2011likelihood},
rewards accuracy in a specific region, $A$\ ($A^{c}$\ denoting the
complement), such as the tail of a predictive distribution for a financial
return. In our numerical work we define $A$ as the lower, and upper tail, of
the predictive distribution, as determined respectively by the 10\% (or 20\%)
and 90\% (or 80\%) quantile of the empirical distribution of $y_{t}$, where we
label these versions of CLS as respectively CLS10 (or CLS20) and CLS90 (or
CLS80). Finally, the score in (\ref{msis}) is the interval score (IS) which is
used in measuring the accuracy of a predictive interval. Here, we use it to
measure the accuracy of $100(1-\alpha)\%$\ predictive interval where
$\alpha=0.05$.

Define the $jth$ scoring rule as $s_{j}(\cdot,\cdot)$, with $s_{j},j\in$ \{LS,
CRPS, CLS10, CLS20, CLS80, CLS90, IS\}, $\hat{\beta}_{j}(y_{1:T})$ as the
optimizer of (\ref{aux_crit}) using $s_{j}$, and $y_{1:T}^{s}(\theta^{i})$ as
the vector of pseudo-data, $y_{1:T}^{s}$, produced using the $ith$ draw
$\theta^{i}$ from the prior $p(\theta)$. For any given rule $s_{j}$, the
summary statistics $\eta\left(  y_{1:T}^{s}\right)  $\textbf{ }and
$\eta\left(  y_{1:T}\right)  $ are defined as the average of the
first-derivative associated with (\ref{aux_crit}) computed, respectively,
using the simulated and observed data, and with both evaluated at $\hat{\beta
}_{j}(y_{1:T})$. That is:%

\begin{equation}
\eta_{j}(y_{1:T}^{s})=\bar{S}_{j}\left\{  y_{1:T}^{s}(\theta^{i});\hat{\beta
}_{j}(y_{1:T})\right\}  , \label{eta_sim}%
\end{equation}
where:%
\[
\bar{S}_{j}\left\{  y_{1:T}^{s}(\theta^{i});\hat{\beta}_{j}(y_{1:T})\right\}
=\left.  T^{-1}\frac{\partial\sum_{t=1}^{T}s_{j}\big(P_{\beta}^{(t)}%
,y_{t+1}^{s}(\theta^{i})\big)}{\partial\beta}\right\vert _{\beta=\hat{\beta
}_{j}(y_{1:T})}%
\]
and:%
\begin{equation}
\eta_{j}(y_{1:T})=\bar{S}_{j}\left\{  y_{1:T};\hat{\beta}_{j}(y_{1:T}%
)\right\}  =0 \label{eta_obs}%
\end{equation}
since%
\[
\bar{S}_{j}\left\{  y_{1:T};\hat{\beta}_{j}(y_{1:T})\right\}  =\left.
T^{-1}\frac{\partial\sum_{t=1}^{T}s_{j}(P_{\beta}^{(t)},y_{t+1})}%
{\partial\beta}\right\vert _{\beta=\hat{\beta}_{j}(y_{1:T})}=0
\]
We take as the distance used in Algorithm 1 the Mahalanobis distance:
\begin{equation}
d\{\eta\left(  y_{1:T}^{s}\right)  \mathbf{,}\eta\left(  y_{1:T}\right)
\}=\sqrt{\left[  \bar{S}_{j}\left\{  y_{1:T}^{s}(\theta^{i});\hat{\beta}%
_{j}(y_{1:T})\right\}  \right]  ^{^{\prime}}\hat{\Sigma}\left[  \bar{S}%
_{j}\left\{  y_{1:T}^{s}(\theta^{i});\hat{\beta}_{j}(y_{1:T})\right\}
\right]  }, \label{dist}%
\end{equation}
where $\hat{\Sigma}$ denotes the inverse of the (estimated) covariance matrix
of $\eta_{j}(y_{1:T}^{s})$ across draws.

The accepted draws of $\theta$ produced by the (modified) Algorithm 1 (see
Footnote 1) are then draws from what we refer to as the loss-based ABC
posterior, $p_{L,\varepsilon}(\theta|\eta(y_{1:T}))$ and, in turn, the ABC
loss-based predictive can be defined as:%
\begin{align}
&  g_{L}(y_{T+1}|y_{1:T})\nonumber\\
&  =\int_{\mathsf{X}}\int_{\mathsf{X}}\int_{{\Theta}}p(y_{T+1}|x_{T+1}%
,\theta\mathbf{,}y_{1:T})p(x_{T+1}|x_{T},\theta)p(x_{T}|\theta,y_{1:T}%
)p_{L,\varepsilon}(\theta|\eta(y_{1:T}))d\theta dx_{T}d{x}_{T+1}.
\label{loss_ABC}%
\end{align}
It is the production and use of $g_{L}(y_{T+1}|y_{1:T})$ that we refer to
hereafter as loss-based ABC prediction (or forecasting) or, for short,
loss-based ABF.

\subsection{Focused Bayesian prediction using the auxiliary model directly}

The use of an auxiliary model with a closed-form predictive to drive the
loss-based ABC prediction exercise does prompt one to explore an obvious
alternative: namely, using the auxiliary model predictive \textit{directly }in
a generalized Bayesian update. Whilst such an approach obviously avoids the
use of a state space specification and, hence, cannot be used to conduct
inference about the parameters of that process, it may well be that for the
purpose of forecasting future observations generated from a state space model,
use of the simpler model in a focused Bayesian up-date may well be adequate.
This is certainly the message gleaned from the numerical work undertaken in
\cite{loaiza2021focused}, and this motivates our exploration of this
alternative to loss-based ABF.

Use of the auxiliary model (with unknown parameter vector $\beta$) as the
predictive model underpinning a focused up-date simply involves defining:%
\begin{equation}
p_{w}(\beta|y_{1:T})=\frac{\exp\left[  wS_{T}\left(  \beta\right)  \right]
p(\beta)}{\int_{\mathcal{B}}\exp\left[  wS_{T}\left(  \beta\right)  \right]
p(\beta)d\beta}, \label{fbp_po}%
\end{equation}
where $S_{T}\left(  \beta\right)  $ is as defined in (\ref{aux_crit}), and $w$
is to be selected in a manner described in Section \ref{sec4}. The one-step
ahead predictive is then constructed as:%
\begin{equation}
p_{FBP}(y_{T+1}|y_{1:T})=\int p_{\beta}(y_{t+1}|y_{1:t})p_{w}(\beta
|y_{1:T})d\beta. \label{fbp_pr}%
\end{equation}
Since this principle of using a simple model to produce focused predictions in
a misspecified setting directly mimics the `focused Bayesian prediction'
approach adopted in \cite{loaiza2021focused}, we use the abbreviation FBP
hereafter to refer to the production and use of $p_{FBP}(y_{T+1}|y_{1:T})$.

\section{Simulation Exercise: Performance of Loss-based ABF}

\label{sec4} In this section, we use an extensive set of simulation
experiments to assess the accuracy of loss-based ABF, under both
correctly-specified and misspecified scenarios. We adopt as the assumed
predictive model an SSM that captures the main stylized features of a
financial return, $y_{t}$, and which is specified in such a way that the exact
(i.e. MCMC-based) predictive is available as a comparator. In the
correctly-specified case -- i.e. when the true DGP matches this assumed model
-- this predictive is the gold standard, to which any alternative would be
compared, including any approximate predictive. In the misspecified case, the
exact predictive no longer assumes this status, and the loss-based ABC
predictives should -- as they are designed to do -- produce the best
out-of-sample performance according to the score criterion on which they are
based, and be superior to the misspecified exact predictive in anything other
than log-score. Comparison is also made -- under both scenarios -- with FBP.
Since the latter is based on a generalized up-date of the parameters of an
auxiliary model that is \textit{always} a simplified version of the assumed
SSM for the return, the FBP results always reflect the impact of misspecification.

\subsection{Simulation design}

\label{sec4.1} Here, the predictive class, $P_{\theta}^{(t)}$, is defined by
an SSM, namely a simple stochastic volatility (SV) model for a continuously
compounded financial return, $y_{t},$ where $\alpha_{t}$ is the log volatility
at time $t$:
\begin{equation}
y_{t}=\mu+e^{\alpha_{t}/2}e_{t}\quad\text{;}\quad e_{t}\sim N(0,1)
\label{eq12}%
\end{equation}%
\begin{equation}
\alpha_{t}=\bar{h}_{\alpha}+\phi(\alpha_{t-1}-\bar{h}_{\alpha})+w_{t}%
\quad\text{;}\quad w_{t}\sim N(0,\sigma_{\alpha}^{2}) \label{eq13}%
\end{equation}

\begin{equation}
\alpha_{1}\sim N\left(  \bar{h}_{\alpha},\frac{\sigma_{\alpha}^{2}}{1-\phi
^{2}}\right)  , \label{eq14}%
\end{equation}
with $\theta=(\phi,\sigma_{\alpha}^{2},\mu,\bar{h}_{\alpha})^{\prime}$, where
$\bar{h}_{\alpha}$ is the marginal mean of the log volatility, $\mu$ is the
marginal mean of the return and $\phi$ is the persistence of volatility. We
adopt two alternative specifications for the true DGP : 1) A model that
matches the SV model in (\ref{eq12})-(\ref{eq14}) (i.e. the assumed predictive
model); and 2) An SV model that better replicates the stylized features of
financial returns data, as used by \cite{loaiza2021focused}:
\begin{equation}
h_{t}=\bar{h}+a(h_{t-1}-\bar{h})+\sigma_{h}\eta_{t} \label{eq15}%
\end{equation}%
\begin{equation}
z_{t}=e^{h_{t}/2}\epsilon_{t} \label{eq16}%
\end{equation}%
\begin{equation}
y_{t}=D^{-1}(F_{z}(z_{t})), \label{eq17}%
\end{equation}
where $\eta_{t}\overset{iid}{\sim}N(0,1)$ and $\epsilon_{t}\overset{iid}{\sim
}N(0,1)$ are independent processes, $\left\{  z_{t}\right\}  _{t=1}^{n}$ is a
latent process with stochastic variance $exp(h_{t})$ and $F_{z}$ is the
implied marginal distribution of $z_{t}$, which is evaluated using simulation.
An inverse distribution function associated with a standardized skewed-normal
distribution $D$, is then used to produce the `observed' return $y_{t}$ as in
(\ref{eq17}), so that the data reflects the usual skewness observed in an
empirical returns distribution in addition to time-varying and autocorrelated
volatility. Clearly, DGP 1) defines a correct-specification setting while DGP
2) defines a misspecification setting.

To generate the summary statistics for obtaining $p_{L,\varepsilon}%
(\theta|\eta(y_{1:T}))$ we adopt two different auxiliary models. It is
necessary to choose an auxiliary model with a closed-form predictive and we
have chosen our auxiliary models to meet this requirement, while still being
reasonable representations of the assumed SSM. The auxiliary models are
defined as follows: (i) a Gaussian autoregressive conditionally
heteroscedastic model of order one (ARCH(1)):
\begin{equation}
y_{t}=\beta_{0}+\sqrt{\alpha_{t}}e_{t}\quad\text{;} \quad e_{t}%
\overset{iid}{\sim}N(0,1)\quad; \quad\alpha_{t}=\beta_{1}+\beta_{2}%
(y_{t-1}-\beta_{0})^{2}, \label{arch}%
\end{equation}
and (ii) a Gaussian generalized (G)ARCH(1,1) model:
\begin{equation}
y_{t}=\beta_{0}+\sqrt{\alpha_{t}}e_{t}\quad\text{;} \quad e_{t}%
\overset{iid}{\sim}N(0,1)\quad; \quad\alpha_{t}=\beta_{1}+\beta_{2}%
\alpha_{t-1}+\beta_{3}(y_{t-1}-\beta_{0})^{2}. \label{garch}%
\end{equation}

We take the following steps in producing the loss-based ABC predictives:

\begin{enumerate}
\item Generate $T$ observations of $y_{t}$ from the true DGP. Here, for the
correct-specification scenario, we generate $T=20000$ observations from
(\ref{eq12})-(\ref{eq14}) using parameter values: $(\phi,\sigma_{\alpha}
,\mu,\bar{h}_{\alpha})=(0.95,0.3,0.0009,-1.3)$. For the misspecification
setting, $T=20000$ observations are generated from (\ref{eq15})-(\ref{eq17})
using the parameter values: $(a,\bar{h},\sigma_{h})=(0.9,-0.4581,0.4173)$,
where $D$ defines the standardized skewed normal distribution with shape
parameter $\gamma=-5$.

\item Use observations $t=1,...,10000$ to evaluate $p_{L,\varepsilon}%
(\theta|\eta(y_{1:T}))$ using ABC. In the implementation of the ABC algorithm,
the quantile used to select the number of draws from the prior ($N$) is
allowed to decline as the sample size $T$ increases in accordance with the
theoretical findings in \cite{frazier2018asymptotic}. We use $q_{T}%
=50T^{-3/2}$, which was also used by \cite{martin2019auxiliary} in their
illustrations. Therefore, with 250 draws retained for the purpose of density
estimation this means that $5000000$ replications are used to produce the results.

\begin{enumerate}
\item Correct-specification setting.\medskip\newline Prior specifications:
$\phi\;\sim\; U(0.5,0.99),\; \sigma_{\alpha}\;\sim\; U(0.05,0.4),\; \mu
\;\sim\; N(0,0.5),\;\bar{h}_{\alpha}\;\sim\; N(-1,1)$ are employed.

\item Misspecification setting. \medskip\newline Prior specifications:
$\phi\;\sim\; U(0.5,0.99),\;\sigma_{\alpha}\;\sim\; U(0.05,0.4),\;\mu\;\sim\;
N(0,1),\;\bar{h}_{\alpha}\;\sim\; N(-3,2)$ are employed.\medskip\newline In
both the correctly-specified and misspecified settings, the summary statistics
and the distance criteria are specified as in (\ref{eta_sim})-(\ref{dist}),
with all seven distinct scoring rules adopted, $s_{j},j\in$ \{LS, CRPS, CLS10,
CLS20, CLS80, CLS90, IS\}. All scoring rules in (\ref{ls})-(\ref{msis}) and
the first derivatives of the score criteria have closed-form solutions for the
two auxiliary models adopted here. \medskip
\end{enumerate}

\item After obtaining $p_{L,\varepsilon}(\theta|\eta(y_{1:T}))$ using the
first $10000$ observations, we hold this posterior fixed, and use draws from
it to construct the one-step-ahead predictives for the remaining $10000$ time
points as given in (\ref{loss_ABC}), using the bootstrap particle filter
(\citealp{gordon1993novel}), with 5000 particles drawn at each step. The
average scores, based on the `observed value', $y_{t}$, are then computed
using $s_{j},j\in$ \{LS, CRPS, CLS10, CLS20, CLS80, CLS90, IS\}. We use the
notation \textquotedblleft ABC-LS\textquotedblright\ to denote loss-based ABF
in the case where the log-score (LS) is used in the criterion in
(\ref{aux_crit}). Corresponding abbreviations are used when (\ref{aux_crit})
is constructed using other scoring rules.
\end{enumerate}

When implementing the simulation design for FBP, we adopt as the predictive
class, $P_{\beta}^{(t)}$, the same auxiliary model which is used for the
comparable loss-based ABC predictive, and produce results under the two DGP
scenarios described above. As highlighted earlier, both scenarios are settings
of misspecification in this case, given the mismatch between the assumed
predictive model underpinning FBP and both true SSMs.

The following steps are undertaken in producing the FBPs:

\begin{enumerate}
\item Use observations $t=1,...,10000$ to construct the FBP posterior in
(\ref{fbp_po}) using seven distinct scoring rules $s_{j},j\in$ \{LS, CRPS,
CLS10, CLS20, CLS80, CLS90, IS\}. For each predictive class, and for each
score update, we produce (after thinning) $M=4000$ posterior draws of
$\theta,\theta^{j},j=1,2,...,M$ using a random walk MH (MCMC) algorithm.

\begin{enumerate}
\item Prior specifications for the Gaussian ARCH(1) model:\newline Parameter
vector: $\beta=(\beta_{0},\beta_{1},\beta_{2})$\newline Prior: $\pi
(\beta)\propto\frac{1}{\beta_{1}}\times\mathit{I}[\beta_{1}>0,\beta_{2}%
\in\lbrack0,1)]$

\item Prior specifications for the Gaussian GARCH(1,1) model:\newline
Parameter vector: $\beta=(\beta_{0},\beta_{1},\beta_{2},\beta_{3})$\newline
Prior: $\pi(\beta)\propto\frac{1}{\beta_{1}}\times\mathit{I}[\beta_{1}%
>0,\beta_{2}\in\lbrack0,1),\beta_{3}\in\lbrack0,1)]\times\mathit{I}(\beta
_{2}+\beta_{3}<1)$\medskip\newline When either (\ref{ls}) or (\ref{cls}) is
used to define (\ref{fbp_po}) a scale of $w=1$ is adopted, which is considered
a natural choice in these cases (see \citealp{loaiza2021focused}). In the case
where $w=1$ and $s(P_{\beta}^{(t)},y_{t+1})=\log p_{\beta}(y_{t+1}|y_{1:t})$,
the update in (\ref{fbp_po}) obviously defaults to the conventional
likelihood-based update of the prior defined over $\beta$, and the resultant
predictive equates to the standard Bayesian predictive for the relevant
auxiliary model. However, if (\ref{crps}) is used to determine the posterior
update in (\ref{fbp_po}), the interpretation of $\exp\left[  wS_{T}\left(
\beta\right)  \right]  $ as a pdf is no longer applicable, and the value of
$w$ must be chosen based on some criterion for weighting $\exp\left[
wS_{T}\left(  \beta\right)  \right]  $ and $\pi(\beta)$. As used in
\cite{loaiza2021focused}, we choose to target a value for $w$ that ensures a
rate of posterior update -- when using CRPS -- that is similar to that of the
update based on LS, by defining%
\[
w=\frac{E_{p(\beta|y_{1:T})}\big[\sum_{t=1}^{T}s_{LS}(P_{\beta}^{t}%
,y_{t+1})\big]}{E_{p(\beta|y_{1:T})}\big[\sum_{t=1}^{T}s_{CRPS}(P_{\beta}%
^{t},y_{t+1})\big]}.
\]
The subscript $p(\beta|y_{1:T})$ indicates that the expectation is taken with
respect to the exact posterior distribution for $\beta$. In practice, an
estimate of $w$ is produced as
\begin{equation}
\hat{w}=\frac{\sum_{j=1}^{J}\big[\sum_{t=1}^{T}s_{LS}(P_{\beta^{(j)}}%
^{t},y_{t+1})\big]}{\sum_{j=1}^{J}\big[\sum_{t=1}^{T}s_{CRPS}(P_{\beta^{(j)}%
}^{t},y_{t+1})\big]}%
\end{equation}
using $J$ draws of $\beta$ from $p(\beta|y_{1:T}),\beta^{j},j=1,2,...,J$%
.\medskip\newline The same methodology is adopted when choosing the value for
$w$ when IS is used in the posterior update, and $w$ is estimated in that case
as
\begin{equation}
\hat{w}=\frac{\sum_{j=1}^{J}\big[\sum_{t=1}^{T}s_{LS}(P_{\beta^{(j)}}%
^{t},y_{t+1})\big]}{\sum_{j=1}^{J}\big[\sum_{t=1}^{T}s_{IS}(P_{\beta^{(j)}%
}^{t},y_{t+1})\big]}.
\end{equation}

\end{enumerate}

\medskip

\item After obtaining the FBP posterior using the first $10000$ observations,
we hold this posterior fixed, and use draws from it to construct the
one-step-ahead predictives for the remaining $10000$ time points. The average
scores, based on the `observed value', $y_{t}$, are then computed using
$s_{j},j\in$ \{LS, CRPS, CLS10, CLS20, CLS80, CLS90, IS\}. We label the method
based on the use of scoring rules in (\ref{ls} )-(\ref{msis}) in the FBP
update by: FBP-LS, FBP-CRPS, FBP-CLS10, FBP-CLS20, FBP-CLS80, FBP-CLS90 and
FBP-IS respectively.
\end{enumerate}

As noted above, we also produce the exact predictive, as a comparator for both
loss-based ABF and FBP. The exact marginal posteriors are produced via a
hybrid Gibbs-MH MCMC algorithm, where we apply the sparse matrix sampling
algorithm of \cite{chan2009mcmc} to sample the states, conditional on $\theta
$, and a standard Gibbs algorithm to sample from the conditional posterior of
$\theta$ given the states. Conditional on the draws of $\theta$ and
$\alpha_{T}$, draws of $\alpha_{T+1}$ are produced directly from
$p(\alpha_{T+1}|\alpha_{T},\theta)$, and the exact predictive is constructed
by averaging the measurement densities over the draws of $\alpha_{T+1}$ and
$\theta$.

\subsection{Simulation results: Correct specification}

\label{sec4.2} In Panel A of Table \ref{tab1} we record the predictive
accuracy of loss-based ABF under correct specification of the SSM. For reasons
of space, we report here the results related to the Gaussian GARCH(1,1)
auxiliary model only, relegating the results for the Gaussian ARCH(1)
auxiliary model to the appendix; noting that the qualitative nature of the
conclusions we draw are the same for both auxiliary models. The rows in Panel
A refer to the scoring rule used to generate the summary statistics in the ABC
algorithm, and the columns record the average out-of-sample loss according to
the score nominated in the column heading.

Given the use of proper scoring rules to define the relevant criterion
function in (\ref{aux_crit}), we would anticipate that, under correct
specification, the optimization process would produce essentially equivalent
values for the summary statistics used in the ABC algorithm and, hence,
(essentially) equivalent draws of $\theta.$ This, in turn, would lead to very
similar loss-based predictives and -- as a consequence -- very similar
out-of-sample performance under all scoring rules, and this is exactly what is
observed in Panel A. Moreover, and as tallies with the asymptotic and
numerical analysis in \cite{frazier2019approximate}, the performance of the
log-score based ABC algorithm -- virtually identical, in turn, to all other
versions of the algorithm, as per the above argument -- is also virtually
equivalent to that of the exact Bayesian method, for all out-of-sample scores.

\begin{table}[H]
\caption{{\protect\footnotesize Predictive accuracy of loss-based ABF (Panel
A) and FBP (Panel B) under correct specification of the SSM. The Gaussian
GARCH(1,1) model is adopted as the auxiliary model. The true DGP is the SV
model in (\ref{eq12})-(\ref{eq14}). The rows in Panel A refer to the scoring
rule used in generating the summary statistics in the ABC algorithm underlying
the loss-based ABF results. The rows in Panel B refer to the scoring rule used
in FBP, while the last row in each panel refers to the exact predictive
results. The columns in each panel refer to the scoring rule used to compute
the out-of-sample average scores. The figures in bold are the largest average
scores according to a given out-of-sample measure. The second largest average
score is in italics.}}%
\label{tab1}%
\subcaption*{Correct model specification} \centering
\par
{\tiny
\begin{tabular}
[c]{lrrrrrrr}\hline\hline
&  &  &  &  &  &  & \\
& \multicolumn{7}{c}{\textbf{Panel A: Loss-based ABF}}\\
&  &  &  &  &  &  & \\
& \multicolumn{7}{c}{Average out-of-sample score}\\\cline{2-8}
&  &  &  &  &  &  & \\
& LS & CLS10 & CLS20 & CLS80 & CLS90 & CRPS & IS\\
\textbf{Scoring rule} &  &  &  &  &  &  & \\\cline{1-1}
&  &  &  &  &  &  & \\
ABC-LS & \textit{-0.8189} & \textit{-0.2956} & -0.4842 & \textit{-0.4875} &
\textit{-0.2985} & \textbf{-0.3211} & -3.0575\\
ABC-CLS10 & -0.8244 & -0.2964 & -0.4857 & -0.4907 & -0.3016 & \textit{-0.3216}
& -3.0946\\
ABC-CLS20 & -0.8193 & \textit{-0.2956} & \textit{-0.4841} & -0.4880 &
-0.2990 & \textbf{-0.3211} & -3.0595\\
ABC-CLS80 & -0.8192 & -0.2957 & \textbf{-0.4840} & -0.4879 & -0.2988 &
\textbf{-0.3211} & -3.0595\\
ABC-CLS90 & -0.8237 & -0.2986 & -0.4875 & -0.4885 & -0.2993 & \textit{-0.3216}
& -3.0932\\
ABC-CRPS & \textbf{-0.8188} & \textbf{-0.2955} & \textit{-0.4841} &
\textbf{-0.4874} & \textbf{-0.2983} & \textbf{-0.3211} & \textbf{-3.0551}\\
ABC-IS & \textbf{-0.8188} & \textbf{-0.2955} & \textbf{-0.4840} & -0.4876 &
-0.2986 & \textbf{-0.3211} & \textit{-3.0559}\\\hline
\textbf{Exact} & -0.8191 & -0.2957 & -0.4842 & -0.4876 & -0.2986 & -0.3211 &
-3.0594\\
&  &  &  &  &  &  &
\end{tabular}
}
\par
{\tiny
\begin{tabular}
[c]{lrrrrrrr}
&  &  &  &  &  &  & \\
& \multicolumn{7}{c}{\textbf{Panel B: FBP}}\\
&  &  &  &  &  &  & \\
& \multicolumn{7}{c}{Average out-of-sample score}\\\cline{2-8}
&  &  &  &  &  &  & \\
& LS & CLS10 & CLS20 & CLS80 & CLS90 & CRPS & IS\\
\textbf{Scoring rule} &  &  &  &  &  &  & \\\cline{1-1}
&  &  &  &  &  &  & \\
FBP-LS & \textit{-0.8557} & -0.3119 & -0.5020 & -0.5009 & -0.3111 &
\textit{-0.3238} & \textit{-3.1525}\\
FBP-CLS10 & -0.8989 & \textbf{-0.3046} & \textbf{-0.4934} & -0.5425 &
-0.3380 & -0.3363 & -3.2309\\
FBP-CLS20 & -0.8838 & \textit{-0.3054} & \textit{-0.4941} & -0.5287 &
-0.3274 & -0.3322 & -3.1963\\
FBP-CLS80 & -0.8862 & -0.3273 & -0.5315 & \textit{-0.4959} & \textit{-0.3061}
& -0.3328 & -3.2051\\
FBP-CLS90 & -0.9154 & -0.3470 & -0.5579 & \textbf{-0.4958} & \textbf{-0.3051}
& -0.3408 & -3.2699\\
FBP-CRPS & -0.8616 & -0.3165 & -0.5057 & -0.5055 & -0.3162 & \textbf{-0.3228}
& -3.1930\\
FBP-IS & \textbf{-0.8551} & -0.3099 & -0.5011 & -0.4997 & -0.3094 & -0.3242 &
\textbf{-3.1434}\\\hline
\textbf{Exact} & -0.8191 & -0.2957 & -0.4842 & -0.4876 & -0.2986 & -0.3211 &
-3.0594\\\hline\hline
\end{tabular}
}\end{table}In summary, when the state space model used in the ABC algorithm
is correctly specified, and contingent on the sample size being sufficiently
large, use of the likelihood function of an auxiliary model to produce the
summary statistics is all that is required to produce predictive results that
are equivalent in accuracy to those produce via an exact Bayesian algorithm,
and according to any measure of out-of-sample accuracy. Focusing on a
particular form of predictive accuracy, via the specification of a different
criterion function within ABC, reaps no additional benefit.

In contrast, if one adopts the approach used in \cite{loaiza2021focused}, and
focuses the predictives \textit{directly }via the use of the Gaussian
GARCH(1,1) model in a generalized Bayesian up-date, the different updates do
lead to different predictives, as is evident from the results recorded in
Panel B of Table \ref{tab1}. In particular, the expected appearance of the
largest average scores in diagonal positions (highlighted in the bold font) is
in evidence overall; that is, using FBP with a particular \textit{focus}
yields the best out-of-sample performance according to that same measure of
predictive accuracy or, using the nomenclature of \cite{martin2022optimal},
yields `coherent' predictions. This is simply the consequence of the fact that
-- by construction -- the auxiliary model is misspecified, and it is under
such misspecification that focusing reaps benefits.

However, and as is critical to note in the current context, one is still much
better off using a correctly specified SSM within an ABC algorithm, despite
the approximate nature of the algorithm, than resorting to an FBP approach
(implemented via an exact MCMC algorithm) using a misspecified predictive
model. That is, all of the average scores in Panel B for the FBP methods are
notably lower than the corresponding scores for the loss-based ABF methods in
Panel A and, in turn, lower than the corresponding results for the exact
predictive. This is really the key conclusion to draw from the numerical
results in Table \ref{tab1}.

\subsection{Simulation results: Misspecification}

\label{sec4.3} Panel A in Table \ref{tab2} documents the predictive accuracy
of loss-based ABF when the SSM used in the ABC algorithm is misspecified. As
anticipated, we now observe differences in predictive accuracy across the
columns, which highlights the fact that the different scoring rules used in
the ABC mechanism to generate the summary statistics are producing different
predictives. Further, we can observe that, with minor deviations, the expected
appearance of bold figures on the main diagonals is in evidence; that is,
using ABC with a particular predictive focus in the auxiliary criterion
function yields the best out-of-sample performance according to that same
measure of predictive accuracy, overall. Moreover, in several cases, despite
the approximate nature of the ABC predictives, the focusing produces more
accurate results out-of-sample -- according to the relevant scoring rule --
than does the exact (and now misspecified) predictive. The most obvious
exception to this statement is the italicized result for ABC-LS in the first
column of Panel A, which is no better than the exact predictive result; this
not being surprising since both predictives are driven by the log score in
this case and superiority of the approximate predictive would not be anticipated.

In Panel B, once again we see the coherence of the FBP results, and more
starkly than was observed in Panel B of Table \ref{tab1}. Given that the
Gaussian GARCH(1,1) model is arguably even more misspecified in this case,
with the true DGP including skewness, this is simply evidence of the fact that
(subject to caveats) focusing reaps more benefits the more extreme is the
misspecification (see \citealp{loaiza2021focused}, and
\citealp{martin2022optimal}). As with the loss-based ABF results, in several
cases focusing with the Gaussian GARCH(1,1) model yields superior
out-of-sample performance than using the exact, but misspecified SSM.

Most importantly, when one compares the results in Panels A and B, the
conclusions are mixed. We observe that FBP is doing better than loss-based ABF
in terms of accuracy in the upper tails. These results relating to FBP are
consistent with the findings by \cite{loaiza2021focused}, produced under the
same simulation scenario, where FBP is found to reap particular benefits in
the upper tail. In contrast, ABC-CRPS and ABC-IS outperform the FBP
counterparts. We also observe that loss-based ABF performs slightly better
than FBP when it is driven by LS and CLS10, but only by a small margin; to
three decimal places the ABC and FBP results based on a CLS20 criterion are
equivalent, in terms of average out-of-sample CLS20 .

\begin{table}[H]
\caption{{\protect\footnotesize Predictive accuracy of loss-based ABF (Panel
A) and FBP (Panel B) under misspecification of the SSM. The Gaussian
GARCH(1,1) model is adopted as the auxiliary model. The true DGP is the SV
model in (\ref{eq15})-(\ref{eq17}), whilst the assumed DGP underlying
loss-based ABF is the SV model in (\ref{eq12})-(\ref{eq14}). The rows in Panel
A refer to the scoring rule used in generating the summary statistics in the
ABC algorithm underlying the loss-based ABF results. The rows in Panel B refer
to the scoring rule used in FBP, while the last row in each panel refers to
the exact (but misspecified) predictive results. The columns in each panel
refer to the scoring rule used to compute the out-of-sample average scores.
The figures in bold are the largest average scores according to a given
out-of-sample measure. The second largest average score is in italics.}}%
\label{tab2}%
\subcaption*{Model misspecification} \centering
\par
{\tiny
\begin{tabular}
[c]{lrrrrrrr}\hline\hline
&  &  &  &  &  &  & \\
& \multicolumn{7}{c}{\textbf{Panel A: Loss-based ABF}}\\
&  &  &  &  &  &  & \\
& \multicolumn{7}{c}{Average out-of-sample score}\\\cline{2-8}
&  &  &  &  &  &  & \\
& LS & CLS10 & CLS20 & CLS80 & CLS90 & CRPS & IS\\
\textbf{Scoring rule} &  &  &  &  &  &  & \\\cline{1-1}
&  &  &  &  &  &  & \\
ABC-LS & \textit{-1.3427} & -0.3586 & \textbf{-0.6173} & -0.4900 & -0.2975 &
\textit{-0.5331} & -4.6333\\
ABC-CLS10 & -1.4117 & \textit{-0.3572} & -0.6327 & -0.5122 & -0.3037 &
-0.5616 & \textit{-4.5813}\\
ABC-CLS20 & -1.3737 & \textbf{-0.3553} & \textit{-0.6202} & -0.5062 &
-0.3084 & -0.5427 & -4.7791\\
ABC-CLS80 & -2.0917 & -0.8118 & -1.2925 & \textbf{-0.4675} & \textit{-0.2822}
& -0.6082 & -10.5000\\
ABC-CLS90 & -2.4259 & -0.8961 & -1.4896 & \textit{-0.4715} & \textbf{-0.2777}
& -0.6509 & -12.7820\\
ABC-CRPS & \textbf{-1.3371} & -0.3629 & -0.6214 & -0.4881 & -0.2998 &
\textbf{-0.5309} & -4.7405\\
ABC-IS & -1.4882 & -0.3657 & -0.6648 & -0.5333 & -0.3057 & -0.6025 &
\textbf{-4.2895}\\\hline
\textbf{Exact} & -1.3343 & -0.3618 & -0.6199 & -0.4882 & -0.3003 & -0.5304 &
-4.7357\\
&  &  &  &  &  &  &
\end{tabular}
}
\par
{\tiny
\begin{tabular}
[c]{lrrrrrrr}
&  &  &  &  &  &  & \\
& \multicolumn{7}{c}{\textbf{Panel B: FBP}}\\
&  &  &  &  &  &  & \\
& \multicolumn{7}{c}{Average out-of-sample score}\\\cline{2-8}
&  &  &  &  &  &  & \\
& LS & CLS10 & CLS20 & CLS80 & CLS90 & CRPS & IS\\
\textbf{Scoring rule} &  &  &  &  &  &  & \\\cline{1-1}
&  &  &  &  &  &  & \\
FBP-LS & \textbf{-1.3471} & -0.3694 & -0.6312 & -0.4840 & -0.2954 &
\textit{-0.5340} & \textit{-4.6822}\\
FBP-CLS10 & -1.3669 & \textbf{-0.3593} & \textit{-0.6243} & -0.5094 &
-0.3212 & -0.5380 & -4.8663\\
FBP-CLS20 & -1.4076 & \textit{-0.3601} & \textbf{-0.6201} & -0.5606 &
-0.3749 & -0.5464 & -5.2801\\
FBP-CLS80 & -2.0718 & -0.9227 & -1.3493 & \textbf{-0.4491} & \textit{-0.2657}
& -0.5888 & -8.9718\\
FBP-CLS90 & -2.5938 & -1.1223 & -1.7430 & \textit{-0.4579} & \textbf{-0.2644}
& -0.6357 & -11.5340\\
FBP-CRPS & \textit{-1.3485} & -0.3786 & -0.6378 & -0.4839 & -0.2973 &
\textbf{-0.5319} & -4.7954\\
FBP-IS & -1.4521 & -0.3734 & -0.6765 & -0.5106 & -0.2943 & -0.5979 &
\textbf{-4.3179}\\\hline
\textbf{Exact} & -1.3343 & -0.3618 & -0.6199 & -0.4882 & -0.3003 & -0.5304 &
-4.7357\\\hline\hline
\end{tabular}
}\end{table}

In summary, whilst focusing within the ABC algorithm can reap benefits
relative to a more crude form of direct up-date via the auxiliary model
itself, this dominance may not be uniform, and is likely to depend on the
extent to which the assumed\textbf{ }SSM is itself misspecified. That said,
there is little to lose from going the ABC route, and the motivation for so
doing may be even stronger once the SSM being entertained is truly
intractable. This is explored as part of the empirical exercise to follow.

\section{Empirical Illustration: SV Model with an Intractable Transition}

\label{sec5}

In this section, we conduct a comparative analysis of the accuracy of the two
predictive methods, loss-based ABF and FBP, in a simple empirical setting
where we forecast the daily returns on the S\&P500 index. Here we adopt an SV
model with $\alpha-$ stable transitions as the assumed predictive model, which
is now a truly intractable SSM, for which exact Bayesian prediction -- even if
deemed to be desirable, given the likelihood of misspecification -- is infeasible.

\subsection{Models and computational details}

\label{sec5.1}

There are several empirical studies in which the non-Gaussian features of
financial returns are captured via the use of $\alpha$-stable processes
(\citealp{carr2003finite}; \citealp{peters2012likelihood};
\citealp{lombardi2009indirect}; \citealp{martin2019auxiliary};
\citealp{frazier2019approximate}). With the $\alpha$-stable process not
possessing a closed-form representation for the density function, the exact
Bayesian predictive is not accessible. ABC-based prediction (and inference),
on the other hand, \textit{is} feasible, given that the $\alpha$-stable
process \textit{can} be simulated via the algorithm proposed in Chambers,
Mallows, and Stuck (1976), and such has been the motivation for the use of ABC
in Peters \textit{et al}., 2012, \cite{martin2019auxiliary} and
\cite{frazier2019approximate}. In this spirit, we adopt here an $\alpha
$-stable process to drive the innovation to the (log) volatility of the
financial return.

Specifically, we define the following model for the continuously compounded
return, $y_{t},$ on the S\&P500 index:
\begin{equation}
y_{t}=e^{h_{t}/2}e_{t}\quad\text{;}\quad e_{t}\sim N(0,1) \label{eq32}%
\end{equation}%
\begin{equation}
h_{t}=\omega+\phi h_{t-1}+\sigma_{h}\eta_{t}\quad\text{;}\quad\eta_{t}%
\sim\mathit{S}(\alpha,-1,0,dt=1), \label{eq33}%
\end{equation}
where $h_{t}$ is the log volatility at time $t$, and $\mathit{S}%
(\alpha,-1,0,dt=1)$ denotes an $\alpha$-stable L\'{e}vy process with location
$\mu=0$, scale $\sigma=1$, tail index $\alpha\in(1,2)$, and skewness parameter
$\gamma=-1$. Clearly, the values of $\alpha$ and $\gamma$ control the degree
of leptokurtosis and skewness in the innovations to the log volatility
process, and we choose to fix the degree of (negative) skewness in the model,
via the specification of the particular value for $\gamma$.

We use daily close-to-close returns data (sourced from Global Financial Data)
from 4 January 2010 to 31 December 2019, comprising 2516 observations. The
most recent 500 observations are reserved for the one-step-ahead predictive
assessments, with the initial set of 2016 observations used to produce the
relevant posterior for the unknown parameters, $\theta=(\omega,\phi
,\sigma_{\eta},\alpha)^{\prime}$, from which draws are taken: ABC draws in the
case of loss-based ABF, and MCMC draws in the case of FBP. That is, as in the
simulation exercise, the posterior is produced only once and not up-dated over
the out-of-sample period.

We implement the loss-based ABF approach in this empirical setting with the
same set of summary statistics and distance criteria as specified in
(\ref{eta_sim})-(\ref{dist}), and using five different scoring rules,
$s_{j},j\in$ \{LS, CLS10, CLS20, CLS80, CLS90\}, to define the relevant sample
criterion in (\ref{aux_crit}). We use the Gaussian GARCH(1,1) model as defined
in (\ref{garch}) as the auxiliary model, and employ uniform priors:
$\omega\;\sim\; U(-1,1),\;\phi\;\sim\; U(0.5,0.99),\;\sigma_{h}\;\sim\;
U(0,0.3),\;\alpha\;\sim\; U(1,2)$. FBP is implemented via direct use of the
same Gaussian GARCH(1,1) model as the assumed predictive model in
(\ref{fbp_po}), with the priors for the parameters of the GARCH model the same
as specified in Section \ref{sec4.1}.

\subsection{Empirical forecasting results}

\label{sec5.2}

In Table \ref{tab3}, we document the empirical predictive results for the two
methods. Panel A records the average out-of-sample scores for the loss-based
ABF method, while Panel B records the corresponding results for FBP.

\begin{table}[H]
\caption{{\protect\footnotesize Predictive accuracy of loss-based ABF (Panel
A) and FBP (Panel B) in an empirical setting.}{\protect\small \ }%
{\protect\footnotesize The Gaussian GARCH(1,1) model is adopted as the
auxiliary model. The assumed DGP underlying loss-based ABF is the SV model in
(\ref{eq32})-(\ref{eq33}).{ The rows in Panel A refer to the scoring rule used
in generating the summary statistics in the ABC algorithm. The rows in Panel B
refer to the scoring rule used in the updating in FBP. The columns in each
panel refer to the out-of-sample measure used to compute the average scores.
The figures in bold are the largest average scores according to a given
out-of-sample measure. The second largest average score is in italics.}}}%
\label{tab3}%
\subcaption*{Assumed predictive model: SV model with $\alpha$-stable errors}
\centering
\par
{\tiny
\begin{tabular}
[c]{lrrrrr}\hline\hline
&  &  &  &  & \\
& \multicolumn{5}{c}{\textbf{Panel A: Loss-based ABF}}\\
&  &  &  &  & \\
& \multicolumn{5}{c}{Average out-of-sample score}\\\cline{2-6}
&  &  &  &  & \\
& LS & CLS10 & CLS20 & CLS80 & CLS90\\
\textbf{Scoring rule} &  &  &  &  & \\\cline{1-1}
&  &  &  &  & \\
ABC-LS & 3.3967 & 0.0268 & 0.2923 & 0.5170 & 0.1523\\
ABC-CLS10 & \textbf{3.4058} & \textbf{0.0391} & \textbf{0.3048} & 0.5122 &
0.1500\\
ABC-CLS20 & \textit{3.4053} & \textit{0.0383} & \textit{0.3041} & 0.5126 &
0.1502\\
ABC-CLS80 & 3.3772 & 0.0052 & 0.2713 & \textbf{0.5192} & \textbf{0.1547}\\
ABC-CLS90 & 3.3849 & 0.0134 & 0.2790 & \textit{0.5191} & \textit{0.1540}\\
&  &  &  &  &
\end{tabular}
}
\par
{\tiny
\begin{tabular}
[c]{lrrrrr}\hline
&  &  &  &  & \\
& \multicolumn{5}{c}{\textbf{Panel B: FBP}}\\
&  &  &  &  & \\
& \multicolumn{5}{c}{Average out-of-sample score}\\\cline{2-6}
&  &  &  &  & \\
& LS & CLS10 & CLS20 & CLS80 & CLS90\\
\textbf{Scoring rule} &  &  &  &  & \\\cline{1-1}
&  &  &  &  & \\
FBP-LS & \textbf{3.3467} & -0.0262 & 0.2406 & 0.4801 & 0.1473\\
FBP-CLS10 & 3.1051 & \textbf{0.0463} & \textbf{0.3096} & 0.1903 & -0.1119\\
FBP-CLS20 & 3.1788 & \textit{0.0417} & \textit{0.3085} & 0.2645 & -0.0430\\
FBP-CLS80 & \textit{3.3293} & -0.0515 & 0.2155 & \textbf{0.4860} &
\textit{0.1531}\\
FBP-CLS90 & 3.1941 & -0.1204 & 0.1042 & \textit{0.4820} & \textbf{0.1543}%
\\\hline\hline
\end{tabular}
}\end{table}

The results show that, apart from minor deviations, the bold figures align as
expected along the main diagonal of each panel. Thus, both predictive
mechanisms are capable of producing coherent predictions overall, in this
empirical setting. That is, focussing the predictive mechanism on the scoring
rule that matters out-of sample -- either via the ABC machinery, or via a
direct up-date using the auxiliary model -- does reap benefits.

However, when it comes to the predictive \textit{superiority} of one approach
over another the findings are once again mixed, mirroring the results in the
(misspecification) simulation exercise in Section \ref{sec4.3}. As a
comparison of the corresponding bolded results in Panels A and B of Table
\ref{tab3} indicates, loss-based ABF dominates in terms of both log-score and
upper-tail based accuracy, whilst FBP outperforms loss-based ABF in terms of
lower tail accuracy. Interestingly, and not surprisingly, the use of the
well-specified SSM in the ABC algorithm leads to the predictive scores in any
one particular column of Panel A being much less diverse than the numbers in
the corresponding column in Panel B, and higher -- sometimes much higher -- in
most cases. That is, the ABC approach, in exploiting a more flexible -- and
arguably less misspecified -- model than does the FBP approach, produces
higher out-of-sample scores, on average, as a consequence. In summary, whilst
the best version of FBP may be better than the best version of loss-based ABF
in the lower tail, \textit{overall }the average scores in Panel A are higher
than those in Panel B, highlighting the importance of factoring in a
well-specified predictive model, and the benefits of using ABC to do so.

\section{Discussion}

\label{sec6}

We have developed a new approach for conducting Bayesian prediction in state
space models (SSMs) that does not rely on correct model specification, and
which accommodates model intractability. Termed loss-based approximate
Bayesian forecasting (loss-based ABF), a posterior is constructed using an
approximate Bayesian computation (ABC) algorithm in which the (vector) summary
statistic is produced by maximizing a criterion function -- defined, in turn,
using a closed-form auxiliary predictive -- that rewards a user-specified
measure of predictive accuracy. The resultant predictive -- by construction --
yields more accuracy out-of-sample when accuracy is assessed using that
particular measure. Two comparators are entertained: exact (but misspecified)
Bayesian prediction (available in a simulation setting at least), and
prediction based on a generalized Bayesian up-date using the auxiliary model
alone. We refer to the latter as focussed Bayesian prediction (FBP), given
that it directly mimics the approach with the same name proposed in
\cite{loaiza2021focused}.

The simulation results reveal that the new approach can yield both coherent
predictions, and predictions that are more accurate than the exact (but
misspecified) MCMC-based predictions, and often (if not uniformly) more
accurate than the FBP results. In an empirical setting, using an intractable
SSM within the ABC algorithm, and for which no exact comparator is available,
loss-based ABF is almost always more accurate than FBP -- exploiting as it
does, an empirically realistic and flexible SSM as part of the ABC algorithm.
However, that said, the very best FBP result -- in terms of accurate
prediction of returns in the lower tail -- is superior to the corresponding
loss-based ABF result.

The overall conclusion that we draw is that there is little to lose from
adopting the ABC-based approach, and much to gain. As is consistent with
previous conclusions about the robustness of predictive accuracy to any
inaccuracy in an ABC posterior (\citealp{frazier2019approximate}), the
approximate nature of the loss-based ABF predictives impinges little on
predictive accuracy. This fact, allied with the ability of the algorithm to
\textit{both }focus on the type of predictive accuracy that matters, and to
accommodate intractable state space specifications, makes it an attractive
proposition. That said, in some cases the more straightward and direct
focussing via the FBP approach may still be entertained, in particular when
the specification of a realistic SSM within an ABC algorithm is deemed to be
challenging; but we would view this option as being the less desirable one, as
a general rule.

As a final point, we note that the proposed ABC procedure can\textit{ }be
viewed as a type of likelihood-free `cutting feedback' approach, whereby the
flow of information from the latent states to the static parameters, which
drive the dynamics of the latent states, is supressed
entirely.\footnote{Cutting feedback is a recently proposed method for
producing robust Bayesian inferences under model misspecification; see
\cite{nott2023bayesian} for a recent review and discussion.} In this context,
we learn about the static parameters via ABC and the specific summaries that
drive predictive performance, without conducting inference on the latent
states; inference on the states is then conducted conditional on the
parameters, and the full set of data, via the bootstrap filter. Thus,
inference about the parameters impacts the states, but not the converse, in
contrast to the usual Bayesian treatment of SSMs. We conjecture that there may
be a link between positive forecast performance in misspecified SSMs, and
cutting the feedback between the states and the static parameters, as is
occurring in the loss-based ABC method; however, we leave a detailed analysis
of this conjecture for future research.

\section{Supplementary Appendix}

\label{sec7}

This supplementary appendix documents the predictive accuracy associated with
the loss-based ABC predictives based on the Gaussian ARCH(1) auxiliary model,
plus the predictive accuracy of FBP in the case where the Gaussian ARCH(1) is
used as the assumed predictive model in the generalized updating. The exact
predictive results are also reproduced here for comparison. All these results
supplement those in Section \ref{sec4} of the main text, and lead to the same
qualitative conclusions as detailed therein.

\begin{table}[H]
\caption{{\protect\footnotesize Predictive accuracy of loss-based ABF (Panel
A) vs FBP (Panel B) under correct specification of the SSM. The Gaussian
ARCH(1) model is adopted as the auxiliary model. The true DGP is the SV model
in (\ref{eq12})-(\ref{eq14}).}{\protect\small \ } {\protect\footnotesize {Sub
panels represent the predictive methodology used. The rows in Panel A refer to
the scoring rule used in generating the summary statistics in the ABC
algorithm underlying the loss-based ABF results. The rows in Panel B refer to
the scoring rule used in FBP, while the last row in each panel refers to the
exact predictive results. The columns in each panel refer to the scoring rule
used to compute the out-of-sample average scores. The figures in bold are the
largest average scores according to a given out-of-sample measure. The second
largest average score is in italics.}}}%
\label{tab4}%
\subcaption*{Correct model specification} \centering
\par
{\tiny
\begin{tabular}
[c]{lrrrrrrr}\hline\hline
&  &  &  &  &  &  & \\
& \multicolumn{7}{c}{\textbf{Panel A: Loss-based ABF}}\\
&  &  &  &  &  &  & \\
& \multicolumn{7}{c}{Average out-of-sample score}\\\cline{2-8}
&  &  &  &  &  &  & \\
& LS & CLS10 & CLS20 & CLS80 & CLS90 & CRPS & IS\\
\textbf{Scoring rule} &  &  &  &  &  &  & \\\cline{1-1}
&  &  &  &  &  &  & \\
ABC-LS & \textit{-0.8193} & -0.2959 & -0.4844 & \textit{-0.4875} &
\textit{-0.2985} & \textit{-0.3212} & -3.0605\\
ABC-CLS10 & -0.8215 & -0.2965 & -0.4855 & -0.4883 & -0.2991 & -0.3214 &
-3.0728\\
ABC-CLS20 & -0.8200 & \textit{-0.2958} & \textit{-0.4842} & -0.4883 &
-0.2990 & -0.3213 & -3.0638\\
ABC-CLS80 & -0.8209 & -0.2960 & \textbf{-0.4841} & -0.4888 & -0.2994 &
-0.3215 & -3.0642\\
ABC-CLS90 & -0.8215 & -0.2960 & \textit{-0.4842} & -0.4892 & -0.2997 &
-0.3215 & -3.0712\\
ABC-CRPS & \textbf{-0.8191} & \textbf{-0.2957} & -0.4844 & \textbf{-0.4874} &
\textbf{-0.2984} & \textbf{-0.3211} & \textbf{-3.0585}\\
ABC-IS & -0.8194 & \textit{-0.2958} & \textbf{-0.4841} & -0.4878 & -0.2987 &
\textit{-0.3212} & \textit{-3.0603}\\\hline
\textbf{Exact} & -0.8191 & -0.2957 & -0.4842 & -0.4876 & -0.2986 & -0.3211 &
-3.0594\\
&  &  &  &  &  &  &
\end{tabular}
}
\par
{\tiny
\begin{tabular}
[c]{lrrrrrrr}
&  &  &  &  &  &  & \\
& \multicolumn{7}{c}{\textbf{Panel B: FBP}}\\
&  &  &  &  &  &  & \\
& \multicolumn{7}{c}{Average out-of-sample score}\\\cline{2-8}
&  &  &  &  &  &  & \\
& LS & CLS10 & CLS20 & CLS80 & CLS90 & CRPS & IS\\
\textbf{Scoring rule} &  &  &  &  &  &  & \\\cline{1-1}
&  &  &  &  &  &  & \\
FBP-LS & \textit{-0.8990} & -0.3226 & -0.5178 & \textit{-0.5146} & -0.3200 &
\textit{-0.3287} & \textbf{-3.2972}\\
FBP-CLS10 & -1.1673 & \textit{-0.3171} & \textit{-0.5124} & -0.7591 &
-0.5407 & -0.4235 & -4.0443\\
FBP-CLS20 & -1.0422 & \textbf{-0.3170} & \textbf{-0.5084} & -0.6453 &
-0.4341 & -0.3710 & -3.6861\\
FBP-CLS80 & -1.0273 & -0.4176 & -0.6327 & \textbf{-0.5110} & \textbf{-0.3175}
& -0.3659 & -3.6335\\
FBP-CLS90 & -1.1538 & -0.5226 & -0.7465 & -0.5165 & \textit{-0.3178} &
-0.4171 & -3.9913\\
FBP-CRPS & \textbf{-0.8972} & -0.3299 & -0.5229 & -0.5166 & -0.3238 &
\textbf{-0.3255} & -3.3509\\
FBP-IS & -0.9065 & -0.3232 & -0.5184 & -0.5188 & -0.3236 & -0.3298 &
\textit{-3.3073}\\\hline
\textbf{Exact} & -0.8191 & -0.2957 & -0.4842 & -0.4876 & -0.2986 & -0.3211 &
-3.0594\\\hline\hline
\end{tabular}
}\end{table}

\begin{table}[H]
\caption{{\protect\footnotesize Predictive accuracy of loss-based ABF (Panel
A) vs FBP (Panel B) under misspecification of the SSM. The Gaussian ARCH(1)
model is adopted as the auxiliary model. The true DGP is the SV model in
(\ref{eq15})-(\ref{eq17}).}{\protect\small \ } {\protect\footnotesize {Sub
panels represent the predictive methodology used. The rows in Panel A refer to
the scoring rule used in generating the summary statistics in the ABC
algorithm underlying the loss-based ABF results. The rows in Panel B refer to
the scoring rule used in FBP, while the last row in each panel refers to the
exact predictive results. The columns in each panel refer to the scoring rule
used to compute the out-of-sample average scores. The figures in bold are the
largest average scores according to a given out-of-sample measure. The second
largest average score is in italics.}}}%
\label{tab5}%
\subcaption*{Model misspecification} \centering
\par
{\tiny
\begin{tabular}
[c]{lrrrrrrr}\hline\hline
&  &  &  &  &  &  & \\
& \multicolumn{7}{c}{\textbf{Panel A: Loss-based ABF}}\\
&  &  &  &  &  &  & \\
& \multicolumn{7}{c}{Average out-of-sample score}\\\cline{2-8}
&  &  &  &  &  &  & \\
& LS & CLS10 & CLS20 & CLS80 & CLS90 & CRPS & IS\\
\textbf{Scoring rule} &  &  &  &  &  &  & \\\cline{1-1}
&  &  &  &  &  &  & \\
ABC-LS & \textit{-1.3433} & \textit{-0.3591} & -0.6190 & -0.4876 & -0.2951 &
\textit{-0.5343} & \textit{-4.6070}\\
ABC-CLS10 & -1.5131 & -0.3883 & -0.6954 & -0.5348 & -0.3095 & -0.6185 &
-4.7425\\
ABC-CLS20 & -1.3610 & \textbf{-0.3545} & \textit{-0.6184} & -0.4975 &
-0.3005 & -0.5411 & -4.6149\\
ABC-CLS80 & -1.6676 & -0.6114 & -0.9617 & \textbf{-0.4596} & \textbf{-0.2742}
& -0.5845 & -8.0744\\
ABC-CLS90 & -1.8580 & -0.6700 & -1.0911 & \textit{-0.4768} & \textit{-0.2808}
& -0.6240 & -9.7305\\
ABC-CRPS & \textbf{-1.3359} & -0.3599 & \textbf{-0.6181} & -0.4911 & -0.3027 &
\textbf{-0.5305} & -4.7400\\
ABC-IS & -1.4926 & -0.3786 & -0.6873 & -0.5290 & -0.3053 & -0.6125 &
\textbf{-4.3467}\\\hline
\textbf{Exact} & -1.3343 & -0.3618 & -0.6199 & -0.4882 & -0.3003 & -0.5304 &
-4.7357\\
&  &  &  &  &  &  &
\end{tabular}
}
\par
{\tiny
\begin{tabular}
[c]{lrrrrrrr}
&  &  &  &  &  &  & \\
& \multicolumn{7}{c}{\textbf{Panel B: FBP}}\\
&  &  &  &  &  &  & \\
& \multicolumn{7}{c}{Average out-of-sample score}\\\cline{2-8}
&  &  &  &  &  &  & \\
& LS & CLS10 & CLS20 & CLS80 & CLS90 & CRPS & IS\\
\textbf{Scoring rule} &  &  &  &  &  &  & \\\cline{1-1}
&  &  &  &  &  &  & \\
FBP-LS & \textbf{-1.3691} & -0.3774 & -0.6436 & -0.4893 & -0.3004 &
\textit{-0.5377} & \textit{-4.8014}\\
FBP-CLS10 & -1.4302 & \textbf{-0.3678} & \textit{-0.6310} & -0.5621 &
-0.3764 & -0.5490 & -5.3863\\
FBP-CLS20 & -1.4781 & \textit{-0.3679} & \textbf{-0.6289} & -0.6130 &
-0.4280 & -0.5680 & -5.7077\\
FBP-CLS80 & -1.9668 & -0.8891 & -1.2627 & \textbf{-0.4564} & \textit{-0.2720}
& -0.5827 & -8.5870\\
FBP-CLS90 & -2.3823 & -1.1207 & -1.6189 & \textit{-0.4620} & \textbf{-0.2713}
& -0.6174 & -10.4630\\
FBP-CRPS & \textit{-1.3699} & -0.3888 & -0.6509 & -0.4882 & -0.3015 &
\textbf{-0.5347} & -4.9309\\
FBP-IS & -1.4797 & -0.3910 & -0.7005 & -0.5177 & -0.3005 & -0.6101 &
\textbf{-4.4381}\\\hline
\textbf{Exact} & -1.3343 & -0.3618 & -0.6199 & -0.4882 & -0.3003 & -0.5304 &
-4.7357\\\hline\hline
\end{tabular}
}\end{table}
\bibliographystyle{apalike}
\bibliography{ref}

\begin{thebibliography}{}

\bibitem[Andrieu et~al., 2011]{andrieu:doucet:holenstein:2010}
Andrieu, C., Doucet, A., and Holenstein, R. (2011).
\newblock Particle {M}arkov chain {M}onte {C}arlo.
\newblock {\em Journal of the Royal Statistical Society: Series B (Statistical
  Methodology)}, 72(2):269--342.
\newblock With discussion.

\bibitem[Biau et~al., 2015]{biau2015new}
Biau, G., C{\'e}rou, F., and Guyader, A. (2015).
\newblock New insights into approximate {B}ayesian computation.
\newblock In {\em Annales de l'IHP Probabilit{\'e}s et statistiques},
  volume~51, pages 376--403.

\bibitem[Bissiri et~al., 2016]{bissiri2016general}
Bissiri, P.~G., Holmes, C.~C., and Walker, S.~G. (2016).
\newblock A general framework for updating belief distributions.
\newblock {\em Journal of the Royal Statistical Society: Series B (Statistical
  Methodology)}, 78(5):1103--1130.

\bibitem[Blei et~al., 2017]{blei2017variational}
Blei, D.~M., Kucukelbir, A., and McAuliffe, J.~D. (2017).
\newblock Variational inference: {A} review for statisticians.
\newblock {\em Journal of the American statistical Association},
  112(518):859--877.

\bibitem[Blum, 2010]{blum2010approximate}
Blum, M.~G. (2010).
\newblock Approximate {B}ayesian computation: {A} nonparametric perspective.
\newblock {\em Journal of the American Statistical Association},
  105(491):1178--1187.

\bibitem[Carr and Wu, 2003]{carr2003finite}
Carr, P. and Wu, L. (2003).
\newblock The finite moment log stable process and option pricing.
\newblock {\em The journal of finance}, 58(2):753--777.

\bibitem[Carter and Kohn, 1994]{carter:kohn:1994}
Carter, C.~K. and Kohn, R. (1994).
\newblock On {G}ibbs sampling for state space models.
\newblock {\em Biometrika}, 81(3):541--553.

\bibitem[Chan and Yu, 2020]{chan2020fast}
Chan, J.~C. and Yu, X. (2020).
\newblock Fast and accurate variational inference for large {B}ayesian {VAR}s
  with stochastic volatility.
\newblock {\em CAMA Working Paper}.

\bibitem[Chan and Jeliazkov, 2009]{chan2009mcmc}
Chan, J. C.-C. and Jeliazkov, I. (2009).
\newblock {MCMC} estimation of restricted covariance matrices.
\newblock {\em Journal of Computational and Graphical Statistics},
  18(2):457--480.

\bibitem[Creel and Kristensen, 2015]{CREEL2015}
Creel, M. and Kristensen, D. (2015).
\newblock {ABC} of {SV}: {L}imited information likelihood inference in
  stochastic volatility jump-diffusion models.
\newblock {\em Journal of Empirical Finance}, 31:85--108.

\bibitem[Dean et~al., 2014]{dean:singh:jasra:peters:2011}
Dean, T.~A., Singh, S.~S., Jasra, A., and Peters, G.~W. (2014).
\newblock Parameter estimation for hidden {M}arkov models with intractable
  likelihoods.
\newblock {\em Scandinavian Journal of Statistics}, 41(4):970--987.

\bibitem[Diks et~al., 2011]{diks2011likelihood}
Diks, C., Panchenko, V., and Van~Dijk, D. (2011).
\newblock Likelihood-based scoring rules for comparing density forecasts in
  tails.
\newblock {\em Journal of Econometrics}, 163(2):215--230.

\bibitem[Drovandi et~al., 2011]{drovandi2011approximate}
Drovandi, C.~C., Pettitt, A.~N., and Faddy, M.~J. (2011).
\newblock Approximate {B}ayesian computation using indirect inference.
\newblock {\em Journal of the Royal Statistical Society: Series C (Applied
  Statistics)}, 60(3):317--337.

\bibitem[Drovandi et~al., 2015]{drovandi2015bayesian}
Drovandi, C.~C., Pettitt, A.~N., and Lee, A. (2015).
\newblock Bayesian indirect inference using a parametric auxiliary model.
\newblock {\em Statistical Science}, 30(1):72--95.

\bibitem[Fearnhead, 2011]{fearnhead2011}
Fearnhead, P. (2011).
\newblock Bayesian inference for time series state space models.
\newblock In Brooks, S., Gelman, A., Jones, G., and Meng, X., editors, {\em
  Handbook of {M}arkov {C}hain {M}onte {C}arlo}, chapter~21, pages 513--530.
  Taylor \& Francis.

\bibitem[Fearnhead and Prangle, 2012]{fearnhead2012constructing}
Fearnhead, P. and Prangle, D. (2012).
\newblock Constructing summary statistics for approximate {B}ayesian
  computation: {S}emi-automatic approximate {B}ayesian computation.
\newblock {\em Journal of the Royal Statistical Society: Series B (Statistical
  Methodology)}, 74(3):419--474.

\bibitem[Flury and Shephard, 2011]{Flury2011}
Flury, T. and Shephard, N. (2011).
\newblock Bayesian inference based only on a simulated likelihood.
\newblock {\em Econometric Theory}, 27:933--956.

\bibitem[Frazier et~al., 2023]{frazier2023variational}
Frazier, D.~T., Loaiza-Maya, R., and Martin, G.~M. (2023).
\newblock Variational {B}ayes in state space models: Inferential and predictive
  accuracy.
\newblock {\em Journal of Computational and Graphical Statistics},
  32(3):793--804.

\bibitem[Frazier et~al., 2021]{frazierloss}
Frazier, D.~T., Loaiza-Maya, R., Martin, G.~M., and Koo, B. (2021).
\newblock Loss-based variational {B}ayes prediction.
\newblock {\em arXiv preprint arXiv:2104.14054}.

\bibitem[Frazier et~al., 2019]{frazier2019approximate}
Frazier, D.~T., Maneesoonthorn, W., Martin, G.~M., and McCabe, B.~P. (2019).
\newblock Approximate {B}ayesian forecasting.
\newblock {\em International Journal of Forecasting}, 35(2):521--539.

\bibitem[Frazier et~al., 2018]{frazier2018asymptotic}
Frazier, D.~T., Martin, G.~M., Robert, C.~P., and Rousseau, J. (2018).
\newblock Asymptotic properties of approximate {B}ayesian computation.
\newblock {\em Biometrika}, 105(3):593--607.

\bibitem[Frazier et~al., 2020]{frazier2020model}
Frazier, D.~T., Robert, C.~P., and Rousseau, J. (2020).
\newblock Model misspecification in approximate {B}ayesian computation:
  {C}onsequences and diagnostics.
\newblock {\em Journal of the Royal Statistical Society: Series B (Statistical
  Methodology)}, 82(2):421--444.

\bibitem[Fr\"{u}hwirth-Schnatter, 1994]{fruhwirth-schnatter:1994}
Fr\"{u}hwirth-Schnatter, S. (1994).
\newblock Data augmentation and dynamic linear models.
\newblock {\em J. Time Ser. Anal.}, 15(2):183--202.

\bibitem[Giordani et~al., 2011]{giordani2011}
Giordani, P., Pitt, M., and Kohn, R. (2011).
\newblock Bayesian inference for time series state space models.
\newblock In Geweke, J., Koop, G., and van Dijk, H., editors, {\em The Oxford
  Handbook of {B}ayesian Econometrics}, chapter~3, pages 61--124. OUP.

\bibitem[Gneiting and Raftery, 2007]{gneiting2007strictly}
Gneiting, T. and Raftery, A.~E. (2007).
\newblock Strictly proper scoring rules, prediction, and estimation.
\newblock {\em Journal of the American statistical Association},
  102(477):359--378.

\bibitem[Gordon et~al., 1993]{gordon1993novel}
Gordon, N.~J., Salmond, D.~J., and Smith, A.~F. (1993).
\newblock Novel approach to nonlinear/non-{G}aussian {B}ayesian state
  estimation.
\newblock In {\em IEE proceedings F (radar and signal processing)}, volume 140,
  pages 107--113. IET.

\bibitem[Jacquier and Polson, 2011]{jacquier2011bayesian}
Jacquier, E. and Polson, N. (2011).
\newblock {B}ayesian methods in finance.
\newblock {\em The Oxford Handbook of {B}ayesian Econometrics}, pages 439--512.
\newblock OUP. Eds. Geweke, J., Koop, G. and van Dijk, H.

\bibitem[Jacquier et~al., 1994]{jacquier94}
Jacquier, R., Polson, N.~G., and Rossi, P.~E. (1994).
\newblock {B}ayesian analysis of stochastic volatility models.
\newblock {\em J. Business and Economic Statistics}, 12(4):371--389.
\newblock With discussion.

\bibitem[Jiang and Tanner, 2008]{jiang2008gibbs}
Jiang, W. and Tanner, M.~A. (2008).
\newblock Gibbs posterior for variable selection in high-dimensional
  classification and data mining.
\newblock {\em The Annals of Statistics}, 36(5):2207--2231.

\bibitem[Joyce and Marjoram, 2008]{joyce2008approximately}
Joyce, P. and Marjoram, P. (2008).
\newblock Approximately sufficient statistics and {B}ayesian computation.
\newblock {\em Statistical applications in genetics and molecular biology},
  7(1).

\bibitem[Kim et~al., 1998]{kim1998svl}
Kim, S., Shephard, N., and Chib, S. (1998).
\newblock Stochastic volatility: {L}ikelihood inference and comparison with
  {ARCH} models.
\newblock {\em The Review of Economic Studies}, 65(3):361--393.

\bibitem[Koop and Korobilis, 2018]{koop2018variational}
Koop, G. and Korobilis, D. (2018).
\newblock Variational {B}ayes inference in high-dimensional time-varying
  parameter models.
\newblock {\em SSRN 3246472}.

\bibitem[Lacoste-Julien et~al., 2011]{lacoste2011approximate}
Lacoste-Julien, S., Husz{\'a}r, F., and Ghahramani, Z. (2011).
\newblock Approximate inference for the loss-calibrated {B}ayesian.
\newblock In {\em Proceedings of the Fourteenth International Conference on
  Artificial Intelligence and Statistics}, pages 416--424. JMLR Workshop and
  Conference Proceedings.

\bibitem[Loaiza-Maya et~al., 2021a]{loaiza2021focused}
Loaiza-Maya, R., Martin, G.~M., and Frazier, D.~T. (2021a).
\newblock Focused {B}ayesian prediction.
\newblock {\em Journal of Applied Econometrics}, 36(5):517--543.

\bibitem[Loaiza-Maya et~al., 2021b]{loaiza2020fast}
Loaiza-Maya, R., Smith, M.~S., Nott, D.~J., and Danaher, P.~J. (2021b).
\newblock Fast and accurate variational inference for models with many latent
  variables.
\newblock {\em Forthcoming. Journal of Econometrics}.

\bibitem[Lombardi and Calzolari, 2009]{lombardi2009indirect}
Lombardi, M.~J. and Calzolari, G. (2009).
\newblock Indirect estimation of $\alpha$-stable stochastic volatility models.
\newblock {\em Computational Statistics \& Data Analysis}, 53(6):2298--2308.

\bibitem[Martin et~al., 2023]{martin2023approximating}
Martin, G.~M., Frazier, D.~T., and Robert, C.~P. (2023).
\newblock Approximating {B}ayes in the 21st century.
\newblock {\em Statistical Science}, 38.
\newblock https://doi.org/10.1214/22-STS875.

\bibitem[Martin et~al., 2022]{martin2022optimal}
Martin, G.~M., Loaiza-Maya, R., Maneesoonthorn, W., Frazier, D.~T., and
  Ram{\'\i}rez-Hassan, A. (2022).
\newblock Optimal probabilistic forecasts: {W}hen do they work?
\newblock {\em International Journal of Forecasting}, 38(1):384--406.

\bibitem[Martin et~al., 2019]{martin2019auxiliary}
Martin, G.~M., McCabe, B.~P., Frazier, D.~T., Maneesoonthorn, W., and Robert,
  C.~P. (2019).
\newblock Auxiliary likelihood-based approximate {B}ayesian computation in
  state space models.
\newblock {\em Journal of Computational and Graphical Statistics},
  28(3):508--522.

\bibitem[Nott et~al., 2023]{nott2023bayesian}
Nott, D.~J., Drovandi, C., and Frazier, D.~T. (2023).
\newblock Bayesian inference for misspecified generative models.
\newblock {\em Annual Review of Statistics and Its Application}, 11.

\bibitem[Peters et~al., 2012]{peters2012likelihood}
Peters, G.~W., Sisson, S.~A., and Fan, Y. (2012).
\newblock Likelihood-free {B}ayesian inference for $\alpha$-stable models.
\newblock {\em Computational Statistics \& Data Analysis}, 56(11):3743--3756.

\bibitem[Pritchard et~al., 1999]{pritchard1999population}
Pritchard, J.~K., Seielstad, M.~T., Perez-Lezaun, A., and Feldman, M.~W.
  (1999).
\newblock Population growth of human {Y} chromosomes: {A} study of {Y}
  chromosome microsatellites.
\newblock {\em Molecular biology and evolution}, 16(12):1791--1798.

\bibitem[Quiroz et~al., 2022]{quiroz2018gaussian}
Quiroz, M., Nott, D.~J., and Kohn, R. (2022).
\newblock Gaussian variational approximation for high-dimensional state space
  models.
\newblock {\em https://arXiv:1801.07873}.
\newblock Forthcoming, Bayesian Analysis.

\bibitem[Strickland et~al., 2006]{STRICKLAND2006}
Strickland, C.~M., Forbes, C.~S., and Martin, G.~M. (2006).
\newblock Bayesian analysis of the stochastic conditional duration model.
\newblock {\em Computational Statistics and Data Analysis}, 50(9):2247--2267.

\bibitem[Stroud et~al., 2003]{stroud2003}
Stroud, J.~R., Müller, P., and Polson, N.~G. (2003).
\newblock Nonlinear state-space models with state-dependent variances.
\newblock {\em Journal of the American Statistical Association},
  98(462):377--386.

\bibitem[Tavar{\'e} et~al., 1997]{tavare1997inferring}
Tavar{\'e}, S., Balding, D.~J., Griffiths, R.~C., and Donnelly, P. (1997).
\newblock Inferring coalescence times from {DNA} sequence data.
\newblock {\em Genetics}, 145(2):505--518.

\bibitem[Tran et~al., 2017]{tran2017variational}
Tran, M.-N., Nott, D.~J., and Kohn, R. (2017).
\newblock Variational {B}ayes with intractable likelihood.
\newblock {\em Journal of Computational and Graphical Statistics},
  26(4):873--882.

\bibitem[Wegmann et~al., 2009]{wegmann2009efficient}
Wegmann, D., Leuenberger, C., and Excoffier, L. (2009).
\newblock Efficient approximate {B}ayesian computation coupled with {M}arkov
  chain {M}onte {C}arlo without likelihood.
\newblock {\em Genetics}, 182(4):1207--1218.

\bibitem[Zhang et~al., 2018]{zhang2018advances}
Zhang, C., B{\"u}tepage, J., Kjellstr{\"o}m, H., and Mandt, S. (2018).
\newblock Advances in variational inference.
\newblock {\em IEEE transactions on pattern analysis and machine intelligence},
  41(8):2008--2026.

\bibitem[Zhang, 2006]{zhang2006information}
Zhang, T. (2006).
\newblock Information-theoretic upper and lower bounds for statistical
  estimation.
\newblock {\em IEEE Transactions on Information Theory}, 52(4):1307--1321.

\end{thebibliography}

\end{document}